\documentclass[reqno]{amsart}
\usepackage[english]{babel}

\usepackage{amsmath,amssymb,amsfonts,amsaddr}
\usepackage{exscale,mathtools,bbm,mathrsfs}
\numberwithin{equation}{section}

\usepackage[T1]{fontenc}
\usepackage{palatino,eulervm}
\usepackage{microtype}

\calclayout

\usepackage{graphicx,xcolor}
\definecolor{links}{rgb}{0,0.3,0}
\usepackage[colorlinks=true,citecolor=blue,urlcolor=links,breaklinks=true]{hyperref}

\usepackage{enumerate,tabularx,booktabs}

\newcommand{\bigo}[1]{\mathcal{O}\left(#1\right)}
\DeclareMathOperator{\Max}{Max}
\DeclareMathOperator{\Min}{Min}
\newcommand{\set}[1]{\left\{ #1 \right\}}
\newcommand{\abs}[1]{\left\lvert #1 \right\rvert}
\newcommand{\dd}[1][]{\mathrm{d}^{#1}}

\newcommand{\mytitle}{Massless scalar Feynman diagrams: five loops and beyond}
\title{\mytitle}
\hypersetup{pdftitle={\mytitle},pdfauthor={David J. Broadhurst}}

\begin{document}

\author{D.~J. Broadhurst}
\address{The Open University, Milton Keynes, England}
\thanks{Work completed on leave of absence at CERN, Geneva.}
\date{December 1985}
\thanks{Note: This attempts to be an accurate reproduction of the preprint \href{http://cds.cern.ch/record/164890}{OUT-4102-18}, based on a scan kindly provided by John Gracey and included as an ancillary file. All transcription errors are to be blamed on \href{mailto:erik.panzer@all-souls.ox.ac.uk}{Erik Panzer}, who will be happy to be alerted about any such misprints in order to correct them so that David's ingenious and amazing work may shine immaculately.}

\begin{abstract}
	Several powerful techniques for evaluating massless scalar Feynman diagrams are developed, viz: the solution of recurrence relations to evaluate diagrams with arbitrary numbers of loops in $n=4-2\omega$ dimensions; the discovery and use of symmetry properties to restrict and compute Taylor series in $\omega$; the reduction of triple sums over Chebyshev polynomials to products of Riemann zeta functions; the exploitation of conformal invariance to avoid four-dimensional Racah coefficients. As an example of the power of these techniques we evaluate \emph{all} of the 216 diagrams, with $5$ loops or less, which give finite contributions of order $1/k^2$ or $1/k^4$ to a propagator of momentum $k$ in massless four-dimensional scalar field theories. Remarkably, only 5 basic numbers are encountered: $\zeta(3)$, $\zeta(5)$, $\zeta(7)$, $\zeta(9)$, and the value of the most symmetrical diagram, which is calculated to $14$ significant figures. It is conceivable that these are the only irrationals appearing in $6$-loop beta functions. En route to these results we uncover and only partially explain many remarkable relations between diagrams.
\end{abstract}
\maketitle

\section{Introduction}
Massless Feynman diagrams present a formidable challenge in the analytical computation of the perturbation expansion of quantum field theories. Yet in recent years the techniques of dimensional regularization \cite{1}, expansions over Gegenbauer polynomials in $x$-space \cite{2}, and integration by parts \cite{3} have yielded new results whose simplicity stands in marked contrast to the labour expended in obtaining them.

The extensions and refinements of these methods, reported in this paper, arose from attempts to elucidate two recent results obtained by the author \cite{4}, using the method of integration by parts of Chetyrkin and Tkachov, hereafter referred to as CT \cite{3}. In ref.\ \cite{4} it was shown how to obtain a result for a class of diagrams involving arbitrary numbers of loops in an arbitrary number of dimensions, by solving recurrence relations which reduce the number of loops.
It was also shown how to improve on the Gegenbauer polynomial $x$-space technique, hereafter referred to as GPXT \cite{2}, by solving recurrence relations which change by one unit the exponent $\alpha$ of a propagator $[1/\ell^2]^{\alpha}$, involving a loop momentum $\ell$.
Sections \ref{sec:2} and \ref{sec:3} generalize these two results significantly and elucidate the first using GPXT. Notable new results include:
a recurrence relation which makes no reference to the dimensionality, $n=4-2\omega$, of spacetime or to the gamma functions arising from one-loop integrations;
a solution to this recurrence relation of remarkable simplicity and generality;
a derivation of the four-dimensional result using GPXT \cite{2}, which exposes the origin of the combinatoric factor in the formula \cite{4}
\begin{equation*}
	I_{\ell} = \binom{2\ell}{\ell} \zeta(2\ell-1)
\end{equation*}
for the $\ell$-loop diagram; the reduction of a more difficult class of $\ell$-loop diagrams to a single infinite sum for the two-loop member;
the invariance of this result under a group of $16$ transformations, $Z_2 \times D_4$, where $D_4$ is the symmetry group of the square;
the use of this invariance to expand diagrams in $\omega$ and to relate diagrams of different topologies;
the use of such relations to obtain $\ell$-loop results involving \emph{products} or Riemann zeta functions.

The author's surprise at how deep these results reach into the loop expansion was tempered by an awareness of the narrowness of their scope in the field of practical calculation involving modest numbers of loops. Accordingly it was determined to tackle a wider class of diagrams up to a finite, and hopefully substantial, number of loops.
Sections \ref{sec:4}--\ref{sec:6} report the methods developed and results achieved in a study of \emph{all} of the $216$ diagrams, with $2$, $3$, $4$ and $5$ loops, which give finite contributions to a two-point function of momentum $k$, in massless four-dimensional scalar field theories, and scale as $1/k^2$ or $1/k^4$.
Remarkably, all of these diagrams can be evaluated, eventually, in terms of $\zeta(3)$, $\zeta(5)$, $\zeta(7)$, $\zeta(9)$ and the value of the most difficult symmetrical $5$-loop diagram, which is computed to $14$ significant figures.
It may be that these are the only irrational numbers that can appear in a $6$-loop beta function. In any case, the contrast between the simplicity of the results obtained and the labour entailed by the extensions of the methods of refs.\ \cite{1,2,3,4} is a glaring one and cries out loudly for a new method which elucidates the long series of surprisingly fortunate circumstances, hereafter referred to as `miracles', which made the present paper possible.
The author believes that the demystification of these miracles is a more urgent challenge than the development of computer algorithms to implement existing methods and earnestly hopes that others will be able to trivialize the simple results obtained here with great difficulty.

The remaining sections of the paper are organised as follows.

Section~\ref{sec:2} is concerned with the diagrams of fig.~\ref{fig:1}, for which partial results were found in \cite{4}. Subsection~\ref{sec:2.1} gives the conventions and definitions which enable one to write and solve the recurrence relation of subsection~\ref{sec:2.2} with transparent economy.
It is no exaggeration to say that the adoption of these conventions and definitions was the crucial step in making progress beyond ref.~\cite{4}. Accordingly, the reader is urged to familiarize himself or herself with them. The elucidation of the four-dimensional result, using GPXT in subsection~\ref{sec:2.3}, was similarly crucial in opening up the calculational possibilities investigated in sections \ref{sec:4}--\ref{sec:6}. The results of section~\ref{sec:2} are discussed in subsection~\ref{sec:2.4}.

Section~\ref{sec:3} is concerned with the diagrams of fig.~\ref{fig:4}, the first of which was partially analyzed in ref.~\cite{4}. Subsection~\ref{sec:3.1} solves a recurrence relation on exponents in two ways, to reveal a hidden symmetry, exploited in subsection~\ref{sec:3.2} to yield a Taylor series in $\omega$ up to the level of $\zeta(7)$, typifying a $5$-loop beta function.
The group of symmetries $Z_2 \times D_4$ of subsection~\ref{sec:3.2} is remarkable. Even more remarkable is the relationship between the diagrams of figs.~\ref{fig:1} and \ref{fig:4}, exposed in subsection~\ref{sec:3.3} and used there to obtain a result for arbitrary loops, involving products of $\zeta$ functions. The results of section~\ref{sec:3} are discussed in subsection~\ref{sec:3.4}.

Section~\ref{sec:4} begins with an introduction to the diagrams of figs.~\ref{fig:6}--\ref{fig:9}, which constitute the challenge to be met in the remainder of the paper. In subsection~\ref{sec:4.1} a rationale for the choice of problem is given, followed by an explanation in subsection~\ref{sec:4.2} of how to relate diagrams by gluing and cutting, and by a discussion in subsection~\ref{sec:4.3} of the angular diagrams generated by GPXT \cite{2}.
(Subsections~\ref{sec:4.2} and \ref{sec:4.3} may be redundant to a reader to whom the captions of figs.~\ref{fig:6}--\ref{fig:9} are transparent.)
Two-loop angular diagrams are studied in subsection~\ref{sec:4.4} and are evaluated in subsection~\ref{sec:4.5}, using a method whose essential simplicity contrasts oddly with the author's failure to find any trace of it in mathematical literature. Subsection~\ref{sec:4.6} explains what remains to be calculated.

Section~\ref{sec:5} demystifies a miracle of section~\ref{sec:4}, using conformal invariance, and shows how a systematic use of this invariance transforms the problem in hand to one which avoids the largely uncharted realm of four-dimensional `Racah coefficients' \cite{2,5,6}.

Section~\ref{sec:6} gives the evaluation of the remaining diagrams, which is achieved with the help of three miracles, revealed in subsections~\ref{sec:6.1} to \ref{sec:6.3}. The final results of sections~\ref{sec:4}--\ref{sec:6} are tabulated and discussed in subsection~\ref{sec:6.4}.

Section~\ref{sec:7} summarizes the conclusions, puzzles, and appeals for further work, of previous sections.

\section{Recurrence relation for all loops}
\label{sec:2}

We first establish a number of conventions and definitions, in terms of which a recurrence relation for arbitrary numbers of loops and dimensions is given without explicit reference to the dimensionality of spacetime or to the gamma function. A solution is given, generalizing our previous result \cite{4}. An alternative derivation of the four-dimensional result is then obtained, using GPXT \cite{2}. The section concludes with a discussion of these results.

\subsection{Conventions and definitions}
\label{sec:2.1}

We work exclusively with Euclidean momenta and normalize the external momentum $k$ of each two-point function by imposing the condition $k^2=1$. The measure for integration over loop momentum $\ell$ is chosen to be $\pi^{-n/2} \int \dd[n] \ell$, in $n$ dimensions, thereby suppressing irrelevant powers of $4\pi$. The generic one-loop diagram, with propagators raised to the powers $\alpha_1$ and $\alpha_2$, is then given by
\begin{equation}\begin{split}
	G(\alpha_1,\alpha_2)
	&\equiv
	\left.%
	\pi^{-n/2} \int \dd[n] \ell \left[ \frac{1}{\ell^2} \right]^{\alpha_1} \left[ \frac{1}{(\ell+k)^2} \right]^{\alpha_2}
	\right|_{k^2=1}
	\\
	&= \frac{\Gamma(\alpha_1+\alpha_2 - n/2)}{\Gamma(\alpha_1)\Gamma(\alpha_2)} B(n/2-\alpha_1,n/2-\alpha_2)
\label{eq:2.1}%
\end{split}\end{equation}
where $\Gamma$ and $B$ are gamma and beta functions. To simplify the dependence on $n=4-2\omega$ we use the mapping \cite{4}
\begin{equation}
	\alpha_i = 1+ \omega(a_i-1)
	\label{eq:2.2}%
\end{equation}
for exponents of propagators. The utility of \eqref{eq:2.2} is apparent when one computes the factor \cite{4}
\begin{equation}
	\Delta(\alpha_1,\alpha_2)
	\equiv
	\frac{\alpha_1 G(\alpha_1+1, \alpha_2) + \alpha_2 G(\alpha_1, \alpha_2+1)}{n-\alpha_1 - \alpha_2 -2}
	\label{eq:2.3}%
\end{equation}
which emerges whenever one uses the triangle rule of CT \cite{3} to eliminate a loop consisting of a triangle with exponents $\alpha_1$, $\alpha_2$ and $1$. Eq.~\ref{eq:2.3} then becomes
\begin{equation}
	\Delta(\alpha_1,\alpha_2)
	= \frac{g_2(a_1,a_2)}{a_1 a_2 \omega^2}
	\label{eq:2.4}%
\end{equation}
where
\begin{equation}
	g_{\ell}(\set{a_i})
	\equiv \frac{\prod_{i=1}^{\ell} R(a_i)}{R(A_{\ell})}
	;\qquad
	A_{\ell} \equiv \sum_{i=1}^{\ell} a_i
	\label{eq:2.5}%
\end{equation}
and
\begin{equation}
	R(a)
	\equiv \frac{\Gamma(1-\omega a)}{\Gamma(1+\omega(a-1))}
	= \frac{1}{R(1-a)}
	\label{eq:2.6}%
\end{equation}
Note that in \eqref{eq:2.4} the dependence on $\omega$ factors out and one encounters only the ratio~\eqref{eq:2.6} of gamma functions, via the function
\begin{equation*}
	g_2(a_1, a_2)
	\equiv \frac{R(a_1) R(a_2)}{R(a_1+a_2)}
\end{equation*}
which is the $\ell=2$ version of the general definition~\eqref{eq:2.5}.
One final definition is needed to expose the beautiful simplicity of the recurrence relation of the next subsection:
\begin{equation}
	P(a)
	\equiv
	\frac{[R(0)]^2}{R(a)R(-a)}
	= P(-a)
	\label{eq:2.7}%
\end{equation}
It will soon be clear that to specialize our results to four dimensions only the expansion of \eqref{eq:2.7} in powers of $\omega$ is needed. For fixed $a$ and sufficiently small $\abs{\omega}$, the standard expansion \cite{7} of $\log \Gamma(1+z)$ gives
\begin{equation}
	\log P(a)
	= \sum_{s=3}^{\infty} \left\{ (a+1)^s - a^s - 1 \right\} \frac{\zeta(s) \omega^s}{s}
	+ (a\rightarrow -a)
	\label{eq:2.8}%
\end{equation}
Notice that $\zeta(2)=\pi^2/6$ does not appear and that when $s$ is even the terms in $a^{s-1} \zeta(s)$ cancel. It is for this reason that, in four dimensions, we will not encounter Riemann zeta functions with even arguments, which would otherwise have led to powers of $\pi^2$.

\subsection{Recurrence relation for all dimensions}
\label{sec:2.2}
\begin{figure}
	\centering
	\includegraphics{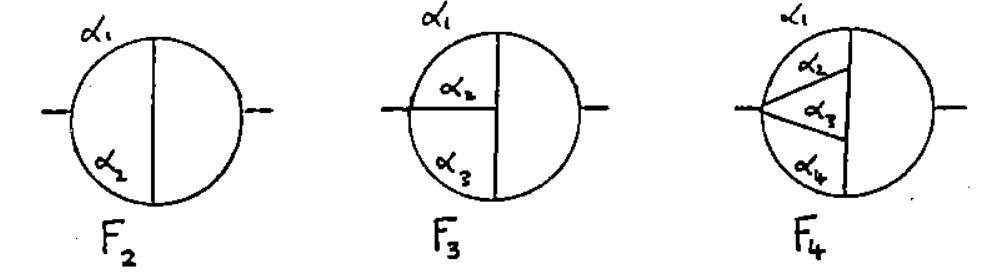}
	\caption{The first three members of the class of diagrams evaluated in eq.~\eqref{eq:2.10}.}%
	\label{fig:1}%
\end{figure}

In \cite{4} we gave, and partially solved, a recurrence relation for the diagrams of fig.~\ref{fig:1}. Here we shall give the complete solution for the $\ell$-loop diagram, $F_{\ell}(\set{\alpha_i})$, when the exponents, $\alpha_1\ldots \alpha_{\ell}$, of the lines connected to the incoming external line are arbitrary. To expose the simplicity of the recurrence relation we express the $\ell$-loop diagram as
\begin{equation*}
	F_{\ell}(\set{\alpha_i})
	= \frac{G(1,1)}{\omega^{2\ell-2}} \frac{g_{\ell}(\set{\alpha_i})}{P(A_{\ell})} f_{\ell}(\set{\alpha_i})
\end{equation*}
where all the terms in the prefactors were defined in the previous subsection and $f_{\ell}$ gives the remaining dependence on the exponents \eqref{eq:2.2}.

Application of the triangle rule of CT \cite{3} then yields the recurrence relation
\begin{equation}\begin{split}
	f_{\ell}(\bullet, a, b, c, d,\bullet)
	&= f_{\ell-1}(\bullet, a, b+c, d, \bullet)/bc \\
	&- f_{\ell-1}(\bullet, a+b, c, d, \bullet)/b(b+c) \\
	&- f_{\ell-1}(\bullet, a, b, c+d, \bullet)/c(b+c)
\label{eq:2.9}%
\end{split}\end{equation}
where $\ell\geq 4$ and the dots indicate the remaining $(\ell-4)$ arguments, which are unchanged. To solve \eqref{eq:2.9} one needs the corresponding results for $\ell=2,3$:
\begin{align*}
	f_2(b,c)
	&= P(b+c)/bc-P(c)/b(b+c) - P(b)/c(b+c) \\
	f_3(a,b,c)
	&= f_2(a,b+c)/bc - f_2 (a+b,c)/b(b+c) - f_2(a,b)/c(b+c)
\end{align*}
These can be thought of as special cases of \eqref{eq:2.9}, with the convention that absent indices are set to zero and that $f_1(a)=P(a)$.

The simplicity of \eqref{eq:2.9} is remarkable: no trace of $\omega$ or $\Gamma$ remains; only addition of arguments and division by sums of arguments is involved.\footnote{%
	The progress we have made in subsection~\ref{sec:2.2}, beyond ref.~\cite{4}, largely results from removing $\Gamma$ functions from the recurrence relation by redefining $f_{\ell}$.
}
It follows that $f_{\ell}(\set{a_i})$ is expressible as a sum of $P$ functions, with arguments given by sums of successive $a_i$ and coefficients which are rational functions of the $a_i$. The full solution is \emph{very} simple:
\begin{equation}
	f_{\ell}
	= \sum_{\ell \geq i > j \geq 0} P(A_i-A_j) 
	\Big/ \prod_{\substack{\ell \geq s \geq 0 \\ s \neq i,j}} (A_i-A_s)(A_s - A_j)
	\label{eq:2.10}%
\end{equation}
where $A_i$ is the sum of the first $i$ arguments and $A_0=0$.
Previously \cite{4} only the term with $i=\ell$ and $j=0$ was found and a restricted result for unit arguments was then obtained by arguing that $f_{\ell} = \bigo{\omega^{2\ell-1}}$ as $\omega \rightarrow 0$. From the full result~\eqref{eq:2.10} this may now be deduced, rather than assumed. Setting $a_i=1$, we readily perform one of the sums in eq.~\eqref{eq:2.10}, obtaining
\begin{equation}
	f_{\ell}(\set{1})
	= \binom{2\ell}{\ell} \sum_{i=1}^{\ell} \frac{(-1)^{\ell+i} i^2}{(\ell+i)!(\ell-i)!} P(i)
	\label{eq:2.11}%
\end{equation}
and taking the limit $\omega\rightarrow 0$, we find
\begin{equation}
	I_{\ell}
	\equiv \lim_{\omega\rightarrow 0} F_{\ell}(\set{1})
	= \binom{2\ell}{\ell} \zeta(2\ell-1)
	\label{eq:2.12}%
\end{equation}
with the aid of expansion~\eqref{eq:2.8}.

The utility of the results~\eqref{eq:2.10}--\eqref{eq:2.12} will be discussed in subsection~\ref{sec:2.4}. First we show how to obtain the four-dimensional result \eqref{eq:2.12} without recourse to dimensional regularization.

\subsection{Recurrence relation for four dimensions}
\label{sec:2.3}

In their paper on the use of Gegenbauer polynomials, Chetyrkin et al.~\cite{2} give a systematic exposition of how to evaluate massless scalar diagrams in $x$-space. In four dimensions the Gegenbauer polynomials are Chebyshev polynomials, which we here define by
\begin{equation}
	C_n(\cos\theta)
	\equiv \frac{\sin n\theta}{\sin \theta};\quad
	n=1,2,\ldots
	\label{eq:2.13}%
\end{equation}
Note that, for later convenience, we label the trivial polynomial, $C_1(x)=1$, by $n=1$, and normalize subsequent polynomials by $C_n(1) = n$. Chebyshev polynomials in momentum space were used to good effect by Rosner \cite{5}, but their applicability appears to be restricted to planar diagrams \cite{2}. In sections~\ref{sec:4}--\ref{sec:6} we shall use them in $x$-space to evaluate a large number of finite four-dimensional planar and non-planar diagrams. Here we use them to derive \eqref{eq:2.12} directly in four dimensions.

In $x$-space the propagator from $x_1$ to $x_2$ is
\begin{equation}
	P_0(x_1,x_2)
	\equiv \frac{1}{(x_1-x_2)^2}
	= \sum_{n=1}^{\infty}
	\left[ x_1^2 x_2^2 R_{12}^n \right]^{-1/2} C_n(\underline{\hat{x}}_1 \cdot \underline{\hat{x}}_2)
	\label{eq:2.14}%
\end{equation}
where $\hat{x}_{1,2}$ are unit four-dimensional Euclidean vectors in the directions of $x_{1,2}$ and
\begin{equation}
	R_{12}
	\equiv \frac{\Max(x_1^2, x_2^2)}{\Min(x_1^2,x_2^2)}
	= R_{21}
	\geq 1
	\label{eq:2.15}%
\end{equation}
is symmetric in $x_{1,2}$ and dimensionless. For our $\ell$-loop result we need to compute the `$\ell$-fold propagator' $P_{\ell}(x_1,x_2)$, defined recursively by
\begin{equation}
	P_{\ell}(x_1,x_2)
	= \pi^{-2} \int \frac{\dd[4] x}{x^2} P_{\ell-1}(x_1,x) P_0(x,x_2)
	\label{eq:2.16}%
\end{equation}
It is straightforward to prove by induction that
\begin{equation}
	P_{\ell}(x_1,x_2)
	= \sum_{i=0}^{\ell} C(\ell,i) \big[ \log R_{12} \big]^i \sum_{n=1}^{\infty} \frac{\left[ x_1^2 x_2^2 R_{12}^n \right]^{-1/2}}{n^{2\ell-i}} C_n(\underline{\hat{x}}_1 \cdot \underline{\hat{x}}_2)
	\label{eq:2.17}%
\end{equation}
using the orthogonality relations\footnote{%
	As in ref.~\cite{2} we normalise the angular measure by $\int \dd \hat{x} = 1$.%
}
\begin{equation}
	\int \dd \underline{\hat{x}} C_n(\underline{\hat{x}}_1 \cdot \underline{\hat{x}}) C_n(\underline{\hat{x}} \cdot \underline{\hat{x}}_2)
	= \frac{1}{n} \delta_{n,m} C_n (\underline{\hat{x}}_1 \cdot \underline{\hat{x}}_2)
	\label{eq:2.18}%
\end{equation}
and splitting the integration over $x^2$ into the three regions
\begin{align*}
	\infty &> x^2 > \Max(x_1^2,x_2^2) \\
	\Max(x_1^2,x_2^2) &> x^2 > \Min(x_1^2, x_2^2) \\
	\Min(x_1^2,x_2^2) &> x^2 > 0
\end{align*}
The coefficients $C(\ell,i)$ in the expansion~\eqref{eq:2.17} are given, recursively, by
\begin{align*}
	i! C(\ell+1,i) &= \sum_{j=i-1}^{\ell} j! C(\ell,j);\quad i=1,2,\ldots,\ell+1 \\
	C(\ell+1,0) &= 2 C(\ell+1,1)
\end{align*}
The solution to these recurrence relations, with $C(0,0)=1$, is
\begin{equation}
	C(\ell,i) = \frac{1}{i!} \binom{2\ell-i}{\ell}
	\label{eq:2.19}%
\end{equation}
which may likewise be proved by induction.
\begin{figure}
	\centering
	\includegraphics{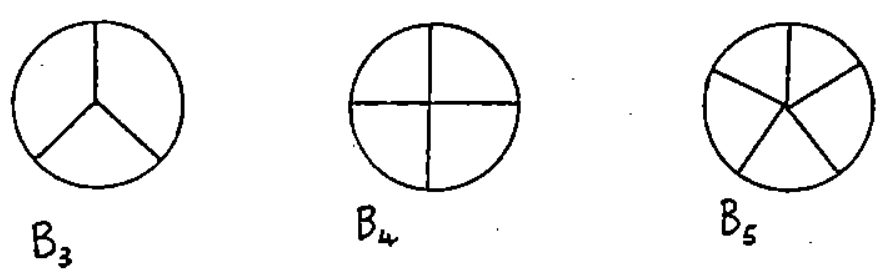}
	\caption{If a spoke of the bubble diagram $B_{\ell+1}$ is cut, one obtains $F_{\ell}$ of fig.~\ref{fig:1}, with unit exponents. If the rim is cut, one obtains $R_{\ell+1}$ of fig.~\ref{fig:4}. In four dimensions $F_{\ell}(\set{1})$ and $R_{\ell+1}(\set{1})$ are both given by eq.~\eqref{eq:2.20}, which gives the coefficient of logarithmic divergence of $B_{\ell+1}$.}%
	\label{fig:2}%
\end{figure}

To prove eq.~\eqref{eq:2.12} we restore the scale factor of $1/k^4$ to the diagrams of fig.~\ref{fig:1}, with unit exponents in four dimensions, and integrate over $k$ to obtain the logarithmically divergent result
\begin{equation*}
	\pi^{-2} \int \frac{\dd[4] k}{k^4} I_{\ell}
\end{equation*}
for the bubble diagram $B_{\ell+1}$ with $\ell+1$ spokes in fig.~\ref{fig:2}. On the other hand we may equally well evaluate $B_{\ell+1}$ in $x$-space, choosing the centre point as origin and obtaining
\begin{equation*}
	\pi^{-2} \int \frac{\dd[4] x}{x^2} P_{\ell}(x,x)
\end{equation*}
Equating the coefficients of logarithmic divergence in $p$- and $x$-space we find that
\begin{equation}
	I_{\ell}
	= x^2 P_{\ell}(x,x)
	= C(\ell,0) \sum_{n=1}^{\infty} \frac{C_n(1)}{n^{2\ell}}
	= \binom{2\ell}{\ell} \zeta(2\ell-1)
	\label{eq:2.20}%
\end{equation}
using eqs.~\eqref{eq:2.13}, \eqref{eq:2.17} and \eqref{eq:2.19}.

It is satisfying to obtain the finite four-dimensional result \eqref{eq:2.20} without recourse to the dimensional regularization required by the triangle rule in the previous subsection.
Eqs.~\eqref{eq:2.13}--\eqref{eq:2.19} will prove very useful in evaluating more complex finite diagrams in later sections.
In general, whenever one could have used the triangle rule in $4-2\omega$ dimensional momentum space to evaluate a finite diagram, it is usually better to use eq.~\eqref{eq:2.17} in four-dimensional $x$-space. 
Note that by the device of relating finite two-point functions to logarithmically divergent bubble diagrams one completely avoids the expansion of $\exp(i k x)$ in terms of Bessel functions, used extensively in ref.~\cite{2}.
No Bessel functions are needed in our subsequent investigations; only the powers and logarithms of eq.~\eqref{eq:2.17} will appear.

\subsection{Discussion of results}
\label{sec:2.4}
\begin{figure}
	\centering
	\includegraphics{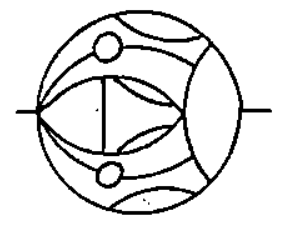}
	\caption{A baroque $13$-loop diagram, easily evaluated in $506-2\epsilon$ dimensions by using eq.~\eqref{eq:2.10} with $\ell=2$ and $\ell=5$.}%
	\label{fig:3}%
\end{figure}

The $n$-dimensional result \eqref{eq:2.10} is amazingly general and frustratingly restrictive.
Its generality consists in giving \emph{all} the diagrams of fig.~\ref{fig:1}, for arbitrary dimensions, loops, and exponents, as a sum of $\ell(\ell+1)/2$ terms, each of which involves the \emph{same} function, with a coefficient and argument given rationally by the dimensionality and the exponents.
All the book-keeping has been done \emph{before} specifying the parameters, making the evaluation of complex diagrams, reducible to fig.~\ref{fig:1}, a routine task in any number of dimensions, with little danger of burdening any symbolic manipulation program.
For example, if one wanted to evaluate fig.~\ref{fig:3} in $506-2\epsilon$ dimensions one would merely have to evaluate $45$ terms, each of which is readily expressible as a ratio of products of gamma functions, of the type $\Gamma(1+m\epsilon)$ with integer $m$, times a rational function of $\epsilon$.
The restrictive nature of the result is equally apparent: unfortunately very few diagrams of interest are reducible to the class of fig.~\ref{fig:1}.

In subsequent sections we shall take some steps towards widening the class of diagrams which can be analytically evaluated with comparative ease.
In section~\ref{sec:3} we consider another, less tractable, infinite class with arbitrary dimensions, loops and exponents.
In section~\ref{sec:4} we evaluate an even harder class to $7$ loops, with unit exponents in four dimensions, and discover a minor miracle, explained in section~\ref{sec:5} by conformal invariance. Then the armoury will be comprehensive enough for a full scale attack on the remaining members of the class of $216$ finite four-dimensional diagrams of section~\ref{sec:6}.

\section{Recurrence relations on exponents and resulting symmetries}
\label{sec:3}
\begin{figure}
	\centering
	\includegraphics{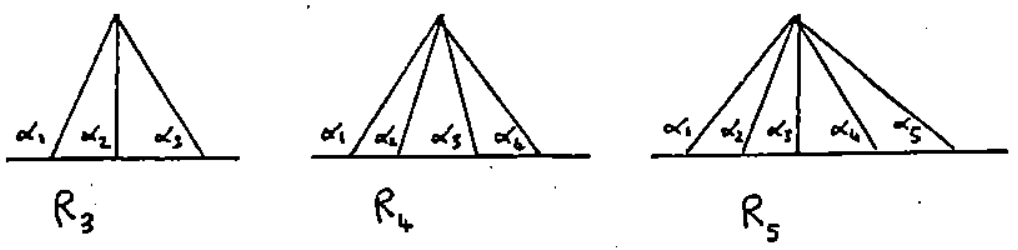}
	\caption{The first three members of the class of diagrams of eq.~\eqref{eq:3.1}.}%
	\label{fig:4}%
\end{figure}

In this section we study the class of diagrams of fig.~\ref{fig:4}, the $\ell$-loop member of which has $\ell+1$ lines connected to the apex of the diagram, with exponents $\alpha_1\ldots \alpha_{\ell+1}$. As in the previous section we seek a method applicable for all dimensions, loops and exponents. There is a simple recurrence relation connecting the $\ell$-loop member, $R_{\ell+1}(\set{\alpha_i})$, to the two-loop member, $R_3(\alpha_1,\alpha_2,\alpha_3)$. Writing\footnote{%
	As before the prefactors have been chosen to simplify the subsequent formulae.%
}
\begin{equation}
	R_{\ell+1}(\set{\alpha_i})
	= \frac{G(1,1)}{\omega^{2\ell-2}} \frac{g_{\ell+1}(\set{a_i})}{R(0)} r_{\ell+1}(\set{a_i})
	\label{eq:3.1}%
\end{equation}
one obtains the same recurrence relation \eqref{eq:2.9} for $r_{\ell}$ as for $f_{\ell}$. But the snag is that $R_3$ (and hence $r_3$) is not believed to be simply expressible in terms of $\Gamma$ functions and hence no general solution like \eqref{eq:2.10} has yet been found. Accordingly we proceed as follows.
First we generalise the method of ref.~\cite{4} to obtain a recurrence relation which relates $R_3(\alpha,\beta,\gamma)$ to $R_3(\alpha,\beta+1,\gamma)$. This is then solved by \emph{two} equivalent single sums on the exponent $\beta$. The result generalizes that found in \cite{4} in the special case $\alpha=\gamma=1$ and is a great improvement on the previous double-sum GPXT result \cite{2} for $R_3$, in two respects.
First the computation of $R_{\ell+1}$, via $R_3$, is greatly simplified, but secondly, and much more significantly, the existence of \emph{two} equivalent expressions for $R_3$ reveals a reflection symmetry, relating $R_3(\alpha,\beta,\gamma)$ and $R_3(\alpha,3n/2-\alpha-\beta-\gamma-2,\gamma)$. When combined with two more symmetries this generates a $16$ member group of symmetries which constrains very tightly the expansion of $R_3$ in powers of $\omega$.
We show that results up to the level of $\zeta(7)$ (relevant to $5$-loop beta functions) can be obtained by simple algebra, without the need for integrations or infinite summations. Moreover we find a result, valid for all loops and involving \emph{products} of Riemann zeta functions, which is perhaps more remarkable than the result \eqref{eq:2.20} relating $I_{\ell}$ to $\zeta(2\ell-1)$.

\subsection{Hidden symmetry}
\label{sec:3.1}

Chetyrkin et al \cite{2} used GPXT to obtain $R_3(\alpha,\beta,\gamma)$ as an infinite double sum, with a summand involving $6$ terms. Here we show how to obtain \emph{two} equivalent single sums, with summands involving $4$ terms. The hidden symmetry thus revealed is the basis for the remainder of section~\ref{sec:3}.

First we use the triangle rule \cite{3} on the left-hand triangle of the first diagram of fig.~\ref{fig:4}. This gives
\begin{multline}
	(n-\alpha-\beta-2) R_3(\alpha,\beta,\gamma) + \alpha R_3(\alpha+1,\beta,\gamma)
	= \\
	G(1,\alpha+\beta+\gamma+1-n/2)
	\Big[ \alpha G(\alpha+1,\beta) + \beta\big\{ G(\alpha,\beta+1) - G(\beta+1,\gamma) \big\} \Big]
	\label{eq:3.2}%
\end{multline}
Next we route the external momentum $k$ through the lines with exponents $\alpha$ and $\gamma$ and operate with $(\partial/\partial k_{\mu})^2$ to obtain
\begin{multline}
	(\alpha+\beta+\gamma+2-n) (\alpha+\beta+\gamma+3-3n/2) R_3(\alpha,\beta,\gamma)
	= -\alpha\gamma R_3(\alpha+1,\beta-1,\gamma+1) \\
	+ (\alpha+\gamma+1-n/2) \big( \alpha R_3(\alpha+1,\beta,\gamma) + \gamma R_3(\alpha,\beta,\gamma+1) \big)
	\label{eq:3.3}%
\end{multline}
It is clear that \eqref{eq:3.2} and \eqref{eq:3.3}, together with the obvious symmetry
\begin{equation}
	R_3(\alpha,\beta,\gamma) = R_3(\gamma,\beta,\alpha)
	\label{eq:3.4}%
\end{equation}
permit one to obtain a recurrence relation on $\beta$ of the form
\begin{align}
	R_3(\alpha,\beta,\gamma)
	&= C(\alpha,\beta,\gamma) H(\alpha,\beta,\gamma)
	\label{eq:3.5} \\
	H(\alpha,\beta,\gamma)
	&= H(\alpha,\beta+1,\gamma) + S(\alpha,\beta,\gamma)
	\label{eq:3.6}%
\end{align}
where $C$ in \eqref{eq:3.5} is chosen to simplify the recurrence relation \eqref{eq:3.6}. Tedious algebra reveals that a suitable form for $C$ is
\begin{equation}\textstyle
	C(\alpha,\beta,\gamma)
	= -2 \frac{
		\Gamma(\frac{n}{2}-1) \Gamma(\alpha+\beta+\gamma+1-n) \Gamma(n-\alpha-\beta-\gamma) \Gamma(\frac{n}{2} - 1 - \beta) \Gamma(n-\beta-2)
	}{
		\Gamma(n-2) \Gamma(\alpha) \Gamma(\gamma) \Gamma(n-\alpha-\beta-1) \Gamma(n-\beta-\gamma-1)
	}
	\label{eq:3.7}%
\end{equation}
With this choice, eqs.~\eqref{eq:3.2}--\eqref{eq:3.7} imply that
\begin{multline}
	S(\alpha,\beta,\gamma)
	= \frac{1}{2} \left[ \frac{
		\Gamma(\gamma) \Gamma(n-2) \Gamma(\frac{n}{2}-\alpha) \Gamma(\alpha+\beta+2-\frac{n}{2}) \Gamma(n-\beta-\gamma-1)
	}{
		\Gamma(\beta+1) \Gamma(\frac{3n}{2} - \alpha - \beta - \gamma -2) \Gamma(\alpha+\beta+\gamma+1-\frac{n}{2}) \Gamma(n-\beta-2)
	}\right]
	\\
	\times \left[ \frac{1}{\alpha+\beta+1-n/2} + \frac{1}{\beta+\gamma+2-n} \right]
	+ (\alpha\leftrightarrow \gamma)
	\label{eq:3.8}%
\end{multline}
In the case $\alpha=\gamma=1$ we recover our previous result \cite{4}.

There are two obvious solutions to \eqref{eq:3.6}:
\begin{subequations}\label{eq:3.9}%
\begin{align}
	H(\alpha,\beta,\gamma)
	&= P(\alpha,\beta,\gamma) - \sum_{i=1}^{\infty} S(\alpha,\beta-i,\gamma)
	\label{eq:3.9a} \\
	H(\alpha,\beta,\gamma)
	&= Q(\alpha,\beta,\gamma) + \sum_{i=0}^{\infty} S(\alpha,\beta+i,\gamma)
	\label{eq:3.9b}%
\end{align}%
\end{subequations}
where $P$ and $Q$ are periodic in $\beta$:
\begin{subequations}\label{eq:3.10}%
\begin{align}
	P(\alpha,\beta,\gamma)
	&= P(\alpha,\beta+1,\gamma)
	\label{eq:3.10a} \\
	Q(\alpha,\beta,\gamma)
	&= Q(\alpha,\beta+1,\gamma)
	\label{eq:3.10b} %
\end{align}
\end{subequations}
Previously we found the results \cite{4}
\begin{subequations}\label{eq:3.11}%
\begin{align}
	P(1,\beta,1)
	&= \pi \cot \pi(\beta+\omega)
	\label{eq:3.11a} \\
	Q(1,\beta,1)
	&= -\pi\cot \pi(\beta+2\omega)
	\label{eq:3.11b} %
\end{align}
\end{subequations}
by careful study of the required cancellation of poles between terms in eq.~\eqref{eq:3.9}. Our job is now to find the corresponding results for general $\alpha$ and $\gamma$. It helps to use the reducible result
\begin{equation}
	R_3(\alpha,0,\gamma) = G(1,\alpha)G(1,\gamma)
	\label{eq:3.12}%
\end{equation}
to establish that
\begin{equation}
	P(\alpha,0,\gamma)
	= \frac{\pi}{2} \big[ \cot \pi (\alpha+\omega) + \cot \pi (\gamma+\omega) \big]
	\label{eq:3.13}%
\end{equation}
Little inspiration is now required to guess the form of $P$ and $Q$ form eqs.~\eqref{eq:3.10}--\eqref{eq:3.13}; the hard work was to cast known results into such simple relations. That being done, the Ans\"{a}tze
\begin{subequations}\label{eq:3.14}%
\begin{align}
	P(\alpha,\beta,\gamma)
	&= \frac{\pi}{2} \big[ \cot \pi (\alpha+\beta+\omega) + \cot \pi (\beta+\gamma+\omega) \big]
	\label{eq:3.14a} \\
	Q(\alpha,\beta,\gamma)
	&= - \frac{\pi}{2} \big[ \cot \pi (\alpha+\beta+2\omega) + \cot \pi (\beta+\gamma+2\omega) \big]
	\label{eq:3.14b} %
\end{align}
\end{subequations}
immediately suggest themselves. We have verified the correctness of eqs.~\eqref{eq:3.14} by numerous high-precision computations of the GPXT double sums \cite{2} and our single sums \eqref{eq:3.9}, for randomly chosen non-integer $\alpha$, $\beta$, $\gamma$ and $\omega$ satisfying the constraints
\begin{align*}
	\alpha+\beta+\gamma &< \Min \left[ 4-3\omega,4-2\omega,5-\omega \right]
	\\
	\abs{\alpha-\gamma} &< 2-3\omega
\end{align*}
which ensure convergence of double and single sums, respectively. It would be more satisfactory to prove eqs.~\eqref{eq:3.14} analytically. Our failure to do so, however, is not surprising when one bears in mind that Chetyrkin et al.~\cite{2} were unable to prove the much more simple result that 
\begin{equation*}
	R_3(\alpha,1,1) = F_2(\alpha,1)
\end{equation*}
starting from their double-sum GPXT result.

Eqs.~\eqref{eq:3.5}, \eqref{eq:3.7}--\eqref{eq:3.9}, and \eqref{eq:3.14} give two equivalent ways of computing $R_3(\alpha,\beta,\gamma)$ as a single sum. The equivalence of these two results implies the reflection symmetry
\begin{equation}
	\frac{R_3(\alpha,\overline{\beta},\gamma)}{R_3(\alpha,\beta,\gamma)}
	=\frac{C(\alpha,\overline{\beta},\gamma)}{C(\alpha,\beta,\gamma)}
	= \frac{R(a+b)R(b+c)}{R(a+b+c-1)R(b+1)}
	\label{eq:3.15}%
\end{equation}
where
\begin{equation}
	\overline{\beta} = 3n/2 - \alpha - \beta - \gamma -2
	\label{eq:3.16}%
\end{equation}
and $\alpha$, $\beta$, $\gamma$ are related to $a$, $b$, $c$ by the mapping \eqref{eq:2.2}.
(The $R$ function on the r.h.s.\ of eq.~\eqref{eq:3.15} is that defined by eq.~\eqref{eq:2.6} in subsection~\ref{sec:2.1}.)
It is ironic that having worked hard to reduce $R_3$ to a single sum we now stumble on eq.~\eqref{eq:3.15}, as if by accident. It will turn out to be more useful than any of the results which spawned it. A more direct derivation could be most illuminating.

In addition to the obvious symmetry \eqref{eq:3.4} and the hidden symmetry \eqref{eq:3.15} there is a third symmetry, whose discovery was intermediate in its difficulty. To uncover it we return to eq.~\eqref{eq:3.1}, which relates $R_{\ell+1}$ to $r_{\ell+1}$, the latter satisfying a recurrence relation identical to that for $f_{\ell}$ in eq.~\eqref{eq:2.9}. In particular we have
\begin{equation}\begin{split}
	r_4(a,b,c,d)
	&= r_3(a,b+c,d)/bc
	\\&
	-r_3 (a+b,c,d)/b(b+c)
	\\&
	-r_3(a,b,c+d)/c(b+c)
	\label{eq:3.17}%
\end{split}\end{equation}
The finiteness of the l.h.s.\ of eq.~\eqref{eq:3.17} when $b=-c$ implies a symmetry property of $r_3$, which can be rearranged to read
\begin{equation}
	r_3(a,b,c) = r_3(a+b,-b,b+c)
	\label{eq:3.18}%
\end{equation}
Our next task is to exploit all three symmetries \eqref{eq:3.4}, \eqref{eq:3.15} and \eqref{eq:3.18} simultaneously.

\subsection{Constrained Taylor series}
\label{sec:3.2}

To exploit the symmetries of $R_3$ we seek a prefactor $N$ in
\begin{equation}
	R_3(\alpha,\beta,\gamma)
	= \frac{G(1,1)}{\omega^2}
	N(a,b,c) r(a,b,c)
	\label{eq:3.19}%
\end{equation}
where $N$ is given simply by $\Gamma$ functions and is chosen such that the three symmetries of $R_3$ emerge in the form
\begin{subequations}\label{eq:3.20}%
\begin{align}
	r(a,b,c)
	&= r(c,b,a)
	\label{eq:3.20a} \\
	&= r(a,1-a-b-c,c)
	\label{eq:3.20b} \\
	&= r(a+b,-b,b+c)
	\label{eq:3.20c} %
\end{align}
\end{subequations}
Note that $r$ must differ from $r_3$, which enjoys the first and third symmetries, but not the second. A little ingenuity was required to arrive at the Ansatz
\begin{equation}
	N(a,b,c) = \big[ g_2(a,b) g_2(b,c) \big]^{1/2} P(b)
	\label{eq:3.21}%
\end{equation}
which is easily verified to lead to eqs.~\eqref{eq:3.20}. The appearance of the square root of $\Gamma$ functions in eq.~\eqref{eq:3.21} is rather novel, but there does not seem to be any other simple way of arriving at eqs.~\eqref{eq:3.20}. This might be taken to suggest that $R_3$, and hence $R_{\ell}$, can be expressed in terms of $\Gamma^{1/2}$. However a counter-example involving the derivatives of $\Gamma$ functions will be given shortly.

The simultaneous symmetries \eqref{eq:3.20} generate a $16$ member group which turns out to be $Z_2 \times D_4$, where $D_4$ is the dihedral group of order $8$, i.e.\ the symmetry group of the square. We now consider how this group of symmetries constrains the expansion of $r$ in powers of $\omega$, needed for the evaluation of diagrams, reducible to the class of fig.~\ref{fig:4}, with divergent subdiagrams. The expansion of $r$ starts with the well known term \cite{2}
\begin{equation*}
	r(a,b,c) = 6 \zeta(3) \omega^3 + \bigo{\omega^4}
\end{equation*}
The term of order $\omega^p$ is a polynomial of order $(p-3)$ in $a$, $b$ and $c$. In general we know the dependence on $a$ and $c$, when $b=1$, since
\begin{equation}
	r(a,1,c) = \frac{(a+c)P(a+c) - aP(a) - cP(c)}{\left[ P(a) P(c) \right]^{1/2} P(a+c) a c (a+c)}
	\label{eq:3.22}%
\end{equation}
is directly obtainable from the result
\begin{equation*}
	R_3(\alpha,1,\gamma)
	= C(\alpha,1,\gamma)\big[ P(\alpha,1,\gamma) - S(\alpha,0,\gamma) \big]
\end{equation*}
which follows from eqs.~\eqref{eq:3.5}--\eqref{eq:3.9} and was previously obtained, in a rather different form, by CT \cite{3}. What then is the most efficient way to incorporate the known two-parameter result \eqref{eq:3.22} into the desired three-parameter expansion of $r$?
Clearly one should expand in terms of the three lowest order invariants of the symmetries \eqref{eq:3.20}, the most complicated of which should be chosen to vanish as $b\rightarrow 1$. The following invariants are particularly convenient:
\begin{align*}
	U &\equiv (a-c)^2 \\
	V &\equiv b^2 + (a+b+c-1)^2-1 \\
	W &\equiv (b^2-1)\left( (a+b+c-1)^2-1 \right)
\end{align*}
They reduce to $(a-c)^2$, $(a+c)^2$ and $0$ in the case $b=1$. Note that $U$ and $V$ are quadratic in $a$, $b$ and $c$, whilst $W$ is quartic. The consequent $16$-fold reduction of the number of terms to be found reflects the underlying group theory.

We take as our expansion
\begin{equation}
	r(a,b,c)
	= \sum_{p=3}^{\infty} \omega^p \sum_{i,j,k} C(i,j,k;p) U^i V^j W^k
	\label{eq:3.23}%
\end{equation}
where the analyticity of $R_3(\alpha,\beta,\gamma)$ in the neighbourhood of $\alpha=\beta=\gamma=1$ and $n=4$ guarantees that
\begin{equation*}
	C(i,j,k;p) = 0
	\qquad\text{for}\qquad
	3+2i+2j+4k > p
\end{equation*}
It follows that the expansion to $\bigo{\omega^6}$ is completely determined by the known result \eqref{eq:3.22}, corresponding to $W=0$. We now show how to obtain the first unknown term, $C(0,0,1;7)$, which requires a knowledge of the term of order $b^4 \omega^7$ in \eqref{eq:3.23}. This can be simply obtained from a knowledge of the $\epsilon^4$ term in 
\begin{equation*}
	\overline{R}_3(\epsilon)
	\equiv \lim_{\omega\rightarrow 0} R_3(1,1+\epsilon,1)
\end{equation*}
which is easily found from our single-sum formulae \eqref{eq:3.5}--\eqref{eq:3.9} and \eqref{eq:3.14}. The result is dramatically simple:
\begin{align}
	\overline{R}_3(\epsilon)
	&= 8 \sum_{s=1}^{\infty} \frac{(-1)^{s+1} s}{(s^2-\epsilon^2)^2}
	\label{eq:3.24} \\
	&= 8 \sum_{p=1}^{\infty} p(1-4^{-p}) \zeta(2p+1) \epsilon^{2p-2}
	\label{eq:3.25}%
\end{align}
where the second form is obtained by expanding the summand of the first. Note that eq.~\eqref{eq:3.24} dashes any remaining hopes that $R_3$ can be expressed solely in terms of $\Gamma$ functions (or their square roots). It shows that even in this very special case $R_3$ involves the second derivative of $\log \Gamma$. But this still leads to a result for $C(0,0,1;7)$ proportional to $\zeta(7)$. Putting together all our previous results we find that
\begin{equation}\label{eq:3.26}\begin{split}
	r(a,b,c) 
	&= r(a,1,c) + \frac{3}{8} (b-1)(a+b+c) \bigg\{ 40\omega^5 \zeta(5)
	\\&
	+20 \omega^6 \left( 5\zeta(6) -2 [\zeta(3)]^2 \right)
	+2 \omega^7 \big( 121 \zeta(7) - 60 \zeta(4) \zeta(3) \big)
	\\&
	+7\omega^7 \zeta(7) \left( 9(b-1)(a+b+c) + 16(a^2+ac+c^2)-6(a+c) \right)
	+\bigo{\omega^8}
	\bigg\}
\end{split}\end{equation}
which means that any diagram reducible to fig.~\ref{fig:4} can be evaluated up to terms involving $\zeta(7)$. In analogy with the known \cite{3} restriction of three- and four-loop beta functions to $\zeta(3)$ and $\zeta(5)$, we expect this to be sufficient for all such diagrams contributing to five-loop beta functions, which should suffice for a while.

It is interesting to compare our route to \eqref{eq:3.26} with that previously taken \cite{4} to arrive at the much less general result
\begin{equation*}
	r(1,b,1) = r(1,1,1) + 15(b-1)(b+2) \omega^5 \zeta(5) + \bigo{\omega^6}
\end{equation*}
Previously we resorted to evaluating single sums as integrals of hypergeometric functions. Here we achieve a much more powerful result by exploiting the hidden symmetry efficiently. The $\zeta$ functions emerge naturally from eqs.~\eqref{eq:2.8} and \eqref{eq:3.25}. Next we show how to use another, more remarkable, symmetry to obtain a result for all loops involving \emph{products} of $\zeta$ functions.

\subsection{Finite diagrams to order $\omega$ for all loops}
\label{sec:3.3}
\begin{figure}
	\centering
	\includegraphics{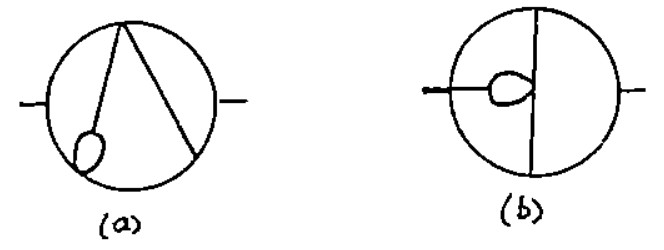}
	\caption{Diagram $(a)$ involves $R_4(1,1+\omega,1,1)$ of fig.~\ref{fig:4}, which is more difficult to evaluate than $F_3(1,1+\omega,1)$ of fig.~\ref{fig:1}, needed for diagram $(b)$. However, eqs.~\eqref{eq:3.30} and \eqref{eq:3.31} enable one to handle diagrams such as $(a)$ very easily, giving eq.~\eqref{eq:3.27} in this particular case.}%
	\label{fig:5}%
\end{figure}

Suppose one needs the singular and finite parts of the four-loop diagram of fig.~\ref{fig:5}a (for example to find a five-loop beta function). This means calculating $R_4(1,1+\omega,1,1)$ to $\bigo{\omega}$, a task which eq.~\eqref{eq:3.26} makes possible, but difficult. By contrast the analogous computation of $F_3(1,1+\omega,1)$, needed for fig.~\ref{fig:5}b, is straightforward.

Here we show how to obtain results such as
\begin{equation}
	R_4(1,1+\omega,1,1)
	= F_3(1,1,1) - 30 [\zeta(3)]^2 \omega + \bigo{\omega^2}
	\label{eq:3.27}%
\end{equation}
for \emph{all} loops and exponents, to $\bigo{\omega}$.

The key observation is that a recurrence relation such as \eqref{eq:2.9} preserves the value of the sum of the arguments. Starting at the $\ell$-loop level the value
\begin{equation*}
	A_{\ell+1} = \sum_{i=1}^{\ell+1} a_i
\end{equation*}
of the sum of the arguments of $r_{\ell+1}$ is preserved in each of the $3^{\ell-2}$ terms generated in working one's way down to the two-loop level of $r_3$. We denote this invariant sum by $S$. A very special symmetry comes into play when $S=0$, since in that case eq.~\eqref{eq:3.20b} shows that $r$, and hence $r_3$, can be evaluated using eq.~\eqref{eq:3.22}. Of course one has to keep track of the prefactors in eqs.~\eqref{eq:3.1} and \eqref{eq:3.19}, which makes $r$ different from $r_3$. But then a miracle happens: the ratio of prefactors is precisely what is required to convert $R_3$ into $F_2$! We find that
\begin{equation}
	R_3(\alpha,\beta,\gamma)
	=F_2(\alpha,\beta)
	=F_2(\gamma,\alpha)
	= F_2(\beta,\gamma)
	\quad\text{for}\quad
	\alpha+\beta+\gamma = 3\left( \frac{n}{2}-1 \right)
	\label{eq:3.28}%
\end{equation}
i.e.\ that the first diagram of fig.~\ref{fig:1} gives the first diagram of fig.~\ref{fig:4} when the sum of the latter's exponents has one particular value. The author's astonishment at this was exceeded by that which greeted the discovery that the recurrence relations for $r_{\ell+1}$ and $f_{\ell}$ preserve such relations for all loops. For example, at the three-loop level
\begin{equation}\label{eq:3.29}\begin{split}
	R_4(\alpha,\beta,\gamma,\delta)
	&=F_3 (\alpha,\beta,\gamma)
	=F_3(\delta,\alpha,\beta)
	\\&
	=F_3(\gamma,\delta,\alpha)
	= F_3(\beta,\gamma,\delta)
	\quad\text{for}\quad
	\alpha+\beta+\gamma+\delta=4\left( \frac{n}{2}-1 \right)
\end{split}\end{equation}
The generalization to all loops is best stated in words: if the sum of the exponents of $R_{\ell+1}$ is $(\ell+1)(\frac{n}{2}-1)$, then $R_{\ell+1}$ can be replaced by $F_{\ell}$ by striking out any one of its arguments and making a cyclic permutation so that the preceding argument is last. This in turn implies an $(\ell+1)$-fold symmetry for $F_{\ell}$ with arbitrary arguments. Note that the relation between $R_{\ell+1}$ and $F_{\ell}$ is between diagrams of quite different topologies.

Relations such as \eqref{eq:3.28} and \eqref{eq:3.29} are fairly trivial when all the exponents are unity and $n=4$, since they degenerate to the known result that \cite{4}
\begin{equation*}
	R_{\ell+1}(\set{1})
	= F_{\ell}(\set{1}) + \bigo{\omega}
	= \binom{2\ell}{\ell} \zeta(2\ell-1) + \bigo{\omega}
\end{equation*}
which follows from cutting the logarithmically divergent bubble diagrams of fig.~\ref{fig:2} in four dimensions \cite{3,4}. That this should generalize to relations with $\ell$ \emph{arbitrary} exponents, in \emph{all} dimensions, was truly unexpected. Again, a more direct proof would be illuminating.

Of course the restriction to $S=0$ is not of practical interest, since diagrams with integer exponents, reducible to the $\ell$-loop diagram of fig.~\ref{fig:4} with non-integer exponents, lead to integer $S>\ell+1$. But the $S=0$ result implies that
\begin{equation}
	R_{\ell+1}(\set{\alpha_i=1+\omega(a_i-1)})
	= F_{\ell}(\set{1}) - \Delta_{\ell} S \omega + \bigo{\omega^2}
	\label{eq:3.30}%
\end{equation}
where $\Delta_{\ell}$ is a pure number (which turns out to be positive).
Note that the arguments of $F_{\ell}$ in \eqref{eq:3.30} can be set to unity at this order of expansion; the dependence on the $a_i$ of $F_{\ell}(\set{\alpha_i=1+\omega(a_i-1)})$ shows up only at $\bigo{\omega^2}$, due to the $(\ell+1)$-fold symmetries of the $\ell$-loop version of relations such as \eqref{eq:3.28} and \eqref{eq:3.29}.
Our task is to find $\Delta_{\ell}$, which will permit the evaluation of singular and finite parts of diagrams with one divergent subdiagram, or the singular parts of diagrams with two. At first sight this would seem difficult, since the terms involving $W$ in the expansion \eqref{eq:3.23} are unknown (except for the first). But in fact they are irrelevant, since
\begin{equation*}
	W \equiv (b^2-1)\left( (a+b+c-1)^2-1 \right)
	=(b^2-1) S(S-2)
\end{equation*}
and however many times the recurrence relation is used to increase $\ell$, such terms will always keep the factor of $S(S-2)$, since $S$ is an invariant. Thus the unknown terms will only show up at $\bigo{\omega^2}$ in \eqref{eq:3.30}. It is hoped that this is enough to convince the reader that $\Delta_{\ell}$ is calculable, in principle, in terms of $\zeta$ functions that appear in the known result \eqref{eq:3.22}, which generates, albeit indirectly, all the $W$-independent terms in the expansion \eqref{eq:3.23}. Suffice it to say that we have devised an efficient algorithm to compute $\Delta_{\ell}$ up to large values of $\ell$, using \texttt{REDUCE3} \cite{8}. The results up to $7$ loops are in Table~\ref{tab:1}.
\begin{table}
	\centering
	\caption{The coefficient $\Delta_{\ell}$ of eq.~\eqref{eq:3.30}, up to $7$ loops.}%
	\label{tab:1}%
	\begin{tabular}{>{$}r<{$}>{$}l<{$}}\toprule
		\ell & \Delta_{\ell}	\\\midrule
		2 & 0 \\
		3 & 6 [\zeta(3)]^2 \\
		4 & 30 \zeta(5) \zeta(3) \\
		5 & 10 [\zeta(5)]^2 + 112 \zeta(7) \zeta(3) \\
		6 & 56 \zeta(7) \zeta(5) + 420 \zeta(9) \zeta(3) \\
		7 & 14 [\zeta(7)]^2 + 240 \zeta(9)\zeta(5) + 1584\zeta(11)\zeta(3)\\\bottomrule
	\end{tabular}
\end{table}

The result for $\Delta_3$ gives eq.~\eqref{eq:3.27} for the four-loop diagram of fig.~\ref{fig:5}a, for which $S=1+2+1+1=5$. The vanishing of $\Delta_2$ is confirmed by inspection of eq.~\eqref{eq:3.26}. The general pattern is clear: all products of two $\zeta$ functions with odd arguments adding to $2\ell$ occur in $\Delta_{\ell}$, with positive integer coefficients. It remains only to give the general rule for these integers:
\begin{equation}
	\Delta_{\ell}
	= \sum_{i=\ell}^{2\ell-3} \left[ 1-(-1)^i \right]
	\left\{ \binom{i+1}{\ell} - \delta_{i,\ell} \right\}
	\zeta(i) \zeta(2\ell-i)
	\label{eq:3.31}%
\end{equation}
This covers all the cases in table~\ref{tab:1} and has been checked to higher loops. Unfortunately an inductive proof has eluded us, owing to the fact that higher order polynomials in $a$ and $c$, generated by the expansion of $P$ in eq.~\eqref{eq:3.22}, are reduced in order by units of $2$ by successive applications of the recurrence relation \eqref{eq:2.9}. Yet again we remark that a direct proof would be illuminating.

\subsection{Discussion of results}
\label{sec:3.4}

The reader may feel, as does the author, that the route to eqs.~\eqref{eq:3.14}, \eqref{eq:3.26} and \eqref{eq:3.31} has been a tortuous and arduous one. Yet the results are not insubstantial: eq.~\eqref{eq:3.14} provides the very simple periodic terms which must be added to the single sums of eq.~\eqref{eq:3.9} to avoid the much more intractable double sums generated by GPXT \cite{2}; 
eq.~\eqref{eq:3.26} allows one to obtain results for diagrams reducible to fig.~\ref{fig:4}, up to the level required for a $5$-loop beta function, without doing any summation or integration at all;
eq.~\eqref{eq:3.31} allows one to calculate, very straightforwardly, such reducible diagrams to all loops, provided they are only modestly divergent. Even so, I am struck by two contrasts in the work reported in this section.
The first contrast is between the specificity of the questions addressed and the generality of the symmetries whose discovery provided the answers: I little thought that in trying to extract the consequences of the obvious starting points of eqs.~\eqref{eq:3.2} and \eqref{eq:3.3} I would be led to the group $Z_2 \times D_4$ underlying eqs.~\eqref{eq:3.20}, or that to obtain results like \eqref{eq:3.27} would entail stumbling on the (to me) amazing $\ell$-loop version of eqs.~\eqref{eq:3.28} and \eqref{eq:3.29}.
The second contrast is between the depth which the results reach in the loop expansion and the narrowness of their scope in the field of practical calculation to modest numbers of loops. I have the strong feeling that anyone who can blur the first contrast will also contribute to removing the second. What is needed is a new approach, which renders trivial the results achieved here by a mixture of notational organization, inspired guesswork and computational exploration.

With such an insight, I believe that breadth of diagrams as well as depth of loops would follow. In the interim this paper will continue by restricting the depth to $5$ loops, but broadening the scope to encompass, eventually, a large number of different types of diagram.

\section{Finite diagrams reducible to triple sums}
\label{sec:4}
\begin{figure}
	\centering
	\includegraphics[width=0.75\textwidth]{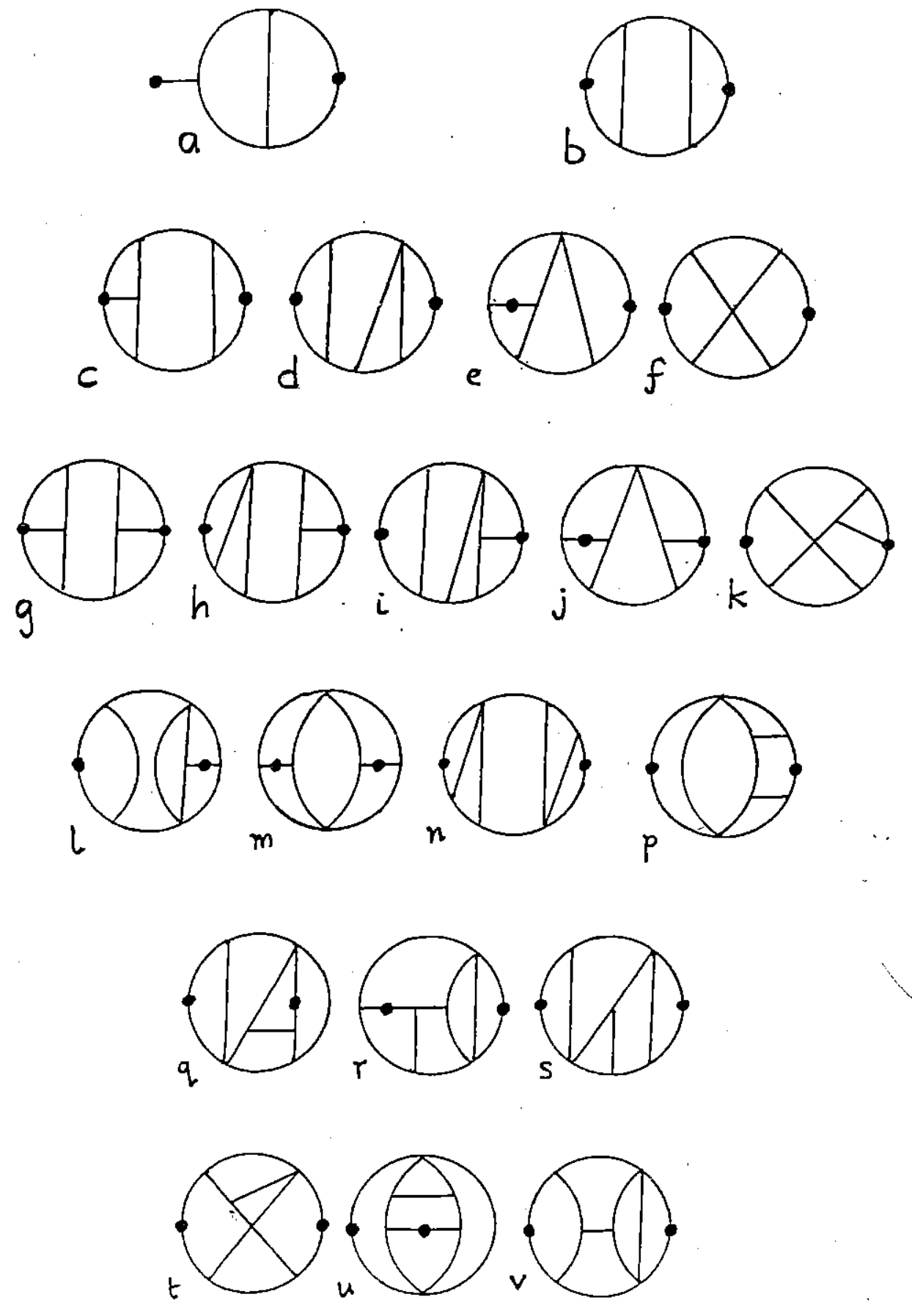}
	\caption{The $21$ diagrams, $a$ to $v$, scale as $1/k^4$ and generate the $216$ diagrams under study by the process of gluing and cutting described in subsection~\ref{sec:4.2}. The external momentum $k$, normalized by $k^2=1$, enters at one blob and leaves at the other. Diagrams $a$, $b$, $c$ and $g$ are given by eq.~\eqref{eq:2.20} with $\ell=2$, $3$, $4$ and $5$. The remaining diagrams of the first three rows are evaluated in section~\ref{sec:4}. Those of the last two rows are shown in section~\ref{sec:5} to be equal to previous diagrams, by exploiting the conformal invariance revealed by the equality of diagrams $e$ and $f$. Diagrams $\ell$ to $p$, on the fourth row, are evaluated in section~\ref{sec:6}. Results for all diagrams are tabulated in subsection~\ref{sec:6.4}.}%
	\label{fig:6}%
\end{figure}

Figs.~\ref{fig:6}--\ref{fig:9} have been carefully designed\footnote{I am grateful to Margaret Broadhurst for her assistance in compiling and organising these figures.} to serve as reference for all the remaining sections. They are introduced here to serve as a guide to what follows and to convince the reader unfamiliar with the use of scissors and glue in four dimensions of their power in remodelling finite diagrams, to be demonstrated shortly. Not all of the information contained in the substantial captions to figs.~\ref{fig:6}--\ref{fig:9} is intended to be studied at this point of the paper; it is intended rather to serve as summary of all that follows.
Suffice it to remark here that fig.~\ref{fig:6} contains a sufficiently complete subset of the $216$ diagrams with which we shall be concerned and that figs.~\ref{fig:7}--\ref{fig:9} give the corresponding `angular diagrams' \cite{2}, which we shall evaluate by GPXT, augmented by novelties and miracles. The $21$ diagrams of fig.~\ref{fig:6} are labelled alphabetically (omitting o, lest it be confused with zero).
The $41$ diagrams of figs.~\ref{fig:7}--\ref{fig:9} are labelled by numerically subscripted letters which enable the reader to cross-reference, without the need of a tabular dictionary.
From henceforward a reference to diagram $e$, for example, will always mean the diagram represented by fig.~\ref{fig:6}e, which is easily located, whilst references to diagrams $e_1$, $e_2$ or $e_3$ will entail a certain amount of searching in figs.~\ref{fig:7}--\ref{fig:9} (which will, on occasions, be lessened by a reference to the appropriate row of one of these figures, e.g.\ to row $2$ of fig.~\ref{fig:7} for diagram $e_1$).

The purpose of section~\ref{sec:4} is to evaluate diagrams $d$, $e$, $h$, $i$ and $j$ of fig.~\ref{fig:6}, via the corresponding angular diagrams of row $2$ of fig.~\ref{fig:7}. After giving a rationale for the choice of diagrams to be evaluated and explaining the use of scissors and glue and the conventions for angular diagrams, we show how to use eqs.~\eqref{eq:2.13}--\eqref{eq:2.19} of subsection~\ref{sec:2.3} to reduce these diagrams to constrained infinite triple sums over rational functions of integers. The resulting mathematical problem does not appear to have been widely studied. We show that a series of `miracles' enables one to use school mathematics to express the diagrams exclusively in terms of $\zeta$ functions. Others are invited to demystify this supernaturally felicitous occurrence.

\subsection{Rationale for the choice of problem}
\label{sec:4.1}

From now on we study only diagrams which are finite in four-dimensional massless scalar field theories and contribute terms proportional to $1/k^2$ or $1/k^4$ to a two-point function of momentum $k$. (Remember that these diagrams correspond to pure numbers, which are positive and have no factors of $4\pi$, because of the conventions given in subsection~\ref{sec:2.1}.)
Table~\ref{tab:2} gives the numbers of such diagrams, to $5$ loops. (It is possible that the table contains the odd overestimate, since it is notoriously difficult to spot the identity of two radically different ways of drawing the same complex diagram. If so, no matter.)
\begin{table}
	\centering
	\caption{Enumeration of diagrams.}%
	\label{tab:2}%
	\begin{tabular}{>{$}r<{$}>{$}r<{$}>{$}r<{$}>{$}r<{$}}
		\text{loops} & 1/k^2 & 1/k^4 & \text{total} \\\midrule
		2 & 1 & 0 & 1\\
		3 & 2 & 2 & 4\\
		4 & 13& 8 & 21\\
		5 & 98&92&190\\ \midrule
		\leq 5 & 114 & 102 & 216\\\bottomrule
	\end{tabular}
\end{table}

The rationale for studying these diagrams is as follows:
\begin{enumerate}[(i)]
	\item
		Finite diagrams are notoriously difficult to obtain by dimensional regularization \cite{3} or blow-by-blow integration over Feynman parameters \cite{9,10}. Finite $\ell$-loop diagrams seem to lead to the nastiest (combinations of) Riemann zeta functions, typically with arguments equal (or summing) to $2\ell-1$ at the $\ell$-loop level. This means that even when one is very lucky (as in the case of eq.~\eqref{eq:2.12}) and obtains the answer by expanding $\Gamma$ functions in $\omega$, after dimensional regularization, the intermediate expressions are of necessity very complex, containing poles of order $1/\omega^{2\ell-1}$ down to $1/\omega$, which will be cancelled by $2\ell-1$ `magic' conspiracies before one encounters $\zeta(2\ell-1)$ (or analogous products) in the finite answer.
		If on the other hand one is foolish enough to resort to integration over Feynman parameters (long since outmoded for massless diagrams \cite{5,2,3}) one can be sure that $\zeta(2\ell-1)$ will emerge only after (at least) $2\ell-1$ integrations of an originally rational function of (at least) $2\ell-1$ variables, and that only the first few of these integrations will be possible without recourse to polylogarithms \cite{11}. For example the author has evaluated all the \emph{massive} two-loop quark-gluon integrals involved in the correlator of a flavour-changing current \cite{9}. The hardest of these (corresponding to masses $m_1$ and $m_2$ on the upper and lower internal lines of diagram $a$) involves integration over four Feynman parameters and gives (eventually)
		\begin{equation}
			I_2(m_1^2/k^2, m_2^2/k^2)
			= F(1) + F(x_1 x_2) - F(x_1) - F(x_2)
			\label{eq:4.1}%
		\end{equation}
		where
		\begin{align*}
			F(x) &\equiv \sum_{n=1}^{\infty} \left[ (2-n\log x)^2 + 2 \right] \frac{x^n}{n^3}
			\\
			x_{1,2} &\equiv m_{1,2}^2 \Big/ \left\{ E+ \left[ E^2-m_1^2 m_2^2 \right]^{1/2} \right\} 
			\\
			E &\equiv (k^2+m_1^2+m_2^2)/2
		\end{align*}
		Attempts to reproduce this have been successful only recently \cite{10} and even so have yielded up a combination of $8$ trilogarithms whose intricate transformation properties \cite{11} obscure the symmetry apparent in \eqref{eq:4.1}. With $m_1=m_2=0$ this is clearly a laborious way of obtaining
		\begin{equation*}
			I_2(0,0) = F(1) = 6 \zeta(3)
		\end{equation*}
		which in 1979 was first seen \cite{12} to be derivable from expanding an exact result of dimensional regularization and is now reduced, by our $\ell$-loop result \cite{4}, to the triviality of computing the number of ways of choosing $2$ out of $4$ objects. These remarks are intended to indicate that, by choosing \emph{finite} diagrams, we are \emph{not} necessarily choosing `to drill many holes where the plank is thinnest', to paraphrase Einstein.

	\item
		The restriction to scalar field theories, by contrast, \emph{is} an admission of defeat in the face of the formidable problem of spin. Modern computer algorithms have trivialised the generation of the polynomials of scalar products of momenta that result from traces and sums over polarisations. Nonetheless spin \emph{is} an essential complication, since beyond the two-loop level these polynomials cannot invariably be expressed in terms of $k^2$ and the squares of loop momenta. At the $\ell$-loop level there are
		\begin{equation*}
			N = (\ell+1)(\ell+2)/2
		\end{equation*}
		scalar products of the $\ell$ loop momenta and the external momentum $k$, which may appear in the numerator of the integrand of a two-point diagram. Unfortunately there are at most
		\begin{equation*}
			D=3\ell
		\end{equation*}
		invariants in the denominator, coming from the $3\ell-1$ propagators of a generic $\ell$-loop $\phi^3$ diagram and a trivial overall power of $1/k^2$. The oversubscription is frightening:
		\begin{equation*}
			\text{deficit} = N-D = (\ell-1)(\ell-2)/2
		\end{equation*}
		It is difficult to overestimate the achievement of CT \cite{3} in surmounting this problem at the three-loop level, where the deficit is merely $10-9=1$, thereby opening the route to the four-loop beta functions of theories with spin.
		We leave it to braver souls to take their tools to the extremely thick plank of `scalarizing' the diagrams involved in the five-loop beta functions of theories with spin. Perhaps the fictitious supersymmetry of recent evaluations of very complex tree graphs \cite{13} has something to offer.

	\item
		The third restriction we impose is that the diagrams scale as $1/k^2$ or $1/k^4$ with the external momentum $k$, i.e.\ that they should not be the product of solely $\phi^3$ couplings at the four- and five-loop levels (which give $1/k^6$ and $1/k^8$) or of one $\phi^4$ coupling and $8$ $\phi^3$ couplings at the five-loop level (which give $1/k^6$). The next two subsections show how this thins our plank considerably, enabling us to throw Bessel functions out of the tool-kit.

	\item
		The final restriction is to $\ell \leq 5$ loops. Rather one might say that we dare to tackle diagrams with $\ell>4$. This was a thicker plank than originally thought to be drillable, since at first sight there seemed no way of avoiding four-dimensional Racah coefficients, which Chetyrkin et al.\ \cite{2} pertinently identified as the main obstacle to deep progress into the loop expansion using GPXT. Fortunately conformal invariance will prove to be the tool that neatly drills through the $\ell=5$ plank, quite dispensing with the need to use Nickel's heroic success in transforming four-dimensional `angular momenta' to three-dimensional ones \cite{6}.
\end{enumerate}
This ends the apologia. Now to work.

\subsection{Scissors and glue}
\label{sec:4.2}

Why are only $21$ out of the $216$ diagrams relevant? The diagrams of fig.~\ref{fig:6} all scale as $1/k^4$. (Since there is no diagram which does so naturally for $\ell=2$ we attach an irrelevant external propagator in diagram $a$, corresponding to trivially multiplying by $1/k^2$). The convention in fig.~\ref{fig:6} is that $k$ flows in at one blob and out at the other. By this notational device we are able unambiguously to depict the non-planar diagrams $e$, $j$, $\ell$, $m$, $q$, $r$ and $u$ without confusion as to which lines do or do not connect.

Now imagine that the diagrams are made of string. Take a pot of glue and stick the two blobs of each diagram together, thereby obtaining $21$ logarithmically divergent bubble diagrams. Manufacture a large number of replicas of each bubble diagram. (If you are by nature cautious it would be as well to have about $50$ replicas of each.)
Now take scissors and cut one line per replica until you have exhausted all the inequivalent ways of cutting each bubble diagram. (Alternatively just cut every line in turn in successive replicas and leave the business of discarding duplicates till later, thereby using up 12 replicas of each of the diagrams $g$--$v$.)
This will produce a large number of finite two-point diagrams, each of which scales as $1/k^2$ (when trimmed of useless external appendages). Now turn your attention to all the vertices of the surviving replicas at which at least $4$ lines meet. Cut out one such vertex per replica and glue together its lines in two groups of at least two lines each. Do this in all inequivalent ways. (Alternatively do it mindlessly and discard duplicates later, thereby using up $35$ of the surviving replicas of the bubble diagram produced by $g$.) This will produce a large number of finite two-point diagrams, each of which scales as $1/k^4$.

It is asserted (with some confidence) that you have now generated \emph{all} the finite diagrams which scale as $1/k^2$ or $1/k^4$ and that there are \emph{no more than} $216$ different ones. It is also probably the case that there are \emph{exactly} $216$ different ones, but no harm will be done if there are fewer. But of course all this gluing and cutting cannot change the coefficient of logarithmic divergence of the corresponding bubble diagrams. So the real substance of our claim is that diagrams $a$--$v$ are convenient ways of labelling the $21$ distinct scalar bubble diagrams, with $3$, $4$, $5$ and $6$ loops, which are logarithmically divergent in four dimensions, containing no subdivergences. Figs.~\ref{fig:7}--\ref{fig:9} will soon be seen to be the source of our confidence that there are \emph{exactly} $21$ of these.

The neglect of contributions to two-point functions scaling as $1/k^6$ or $1/k^8$ is now seen to be crucial. To produce a logarithmically divergent bubble diagram from such diagrams one would need to multiply by $k^2$ or $k^4$ before gluing, thereby creating a numerator impossible to obtain in scalar field theories.

The task is now to find the $21$ pure numbers which give diagrams $a$--$v$, or equivalently to find the coefficients of the logarithmic divergence in the scalar bubble diagrams which diagrams $a$--$v$ generate. The reader interested only in results should turn to table~\ref{tab:3} of subsection~\ref{sec:6.4}, without further ado. What follows now is the method.

\subsection{Angular diagrams}
\label{sec:4.3}
\begin{figure}
	\centering
	\includegraphics[width=0.7\textwidth]{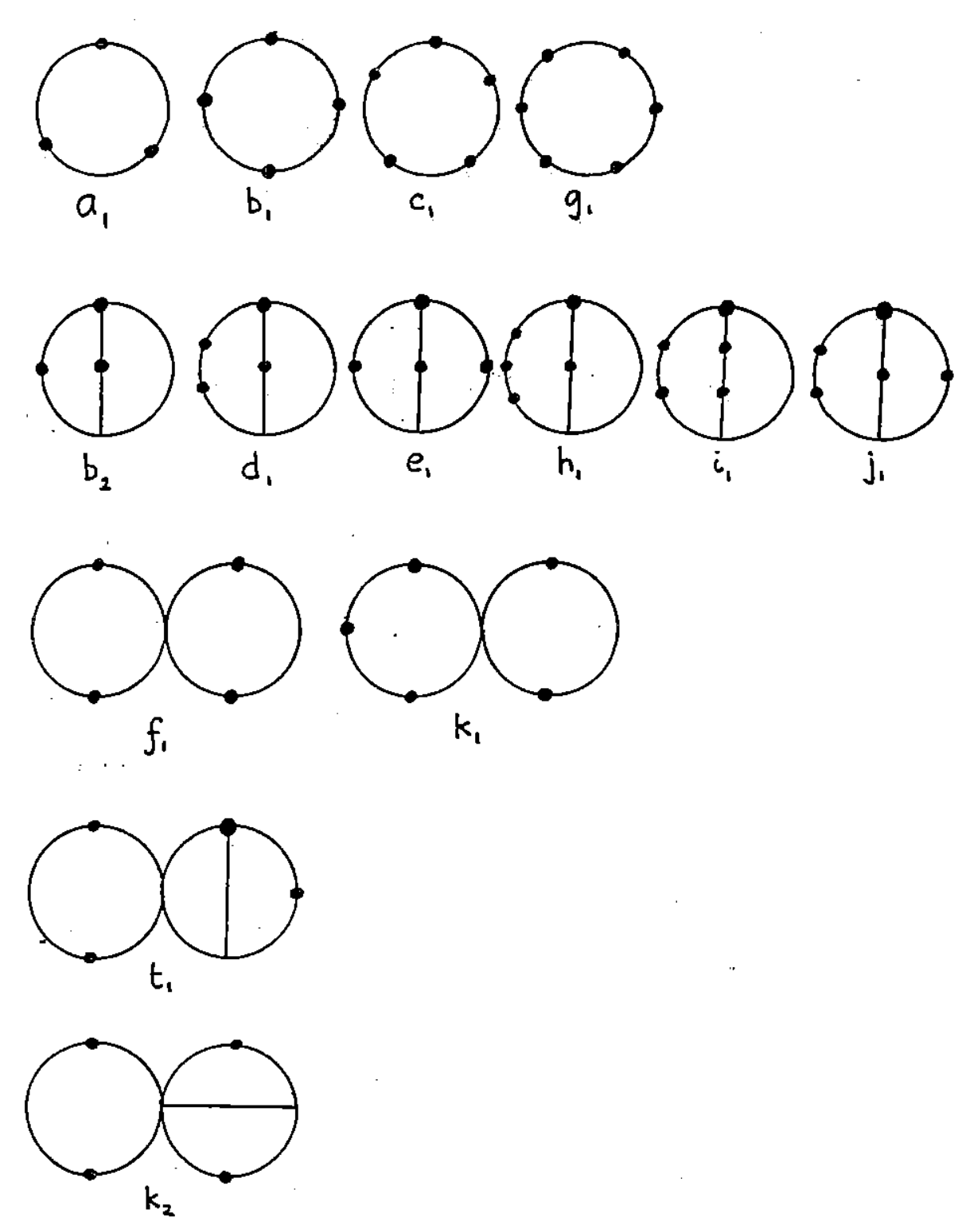}
	\caption{The easiest bubble diagrams. Each row begins with an angular diagram \cite{2} furnished with the minimum number of dots to transform it into a logarithmically divergent bubble diagram. A dot (whether on a line or at a vertex) represents a point joined to the origin (not shown) by a line (not shown). By adding more dots, up to the maximum allowed for $6$-loop bubble diagrams, all bubble diagrams are generated in figs.~\ref{fig:7}--\ref{fig:9}. Those shown here are the easiest, since the first row is given by eq.~\eqref{eq:2.20} and the second by eq.~\eqref{eq:4.6} (with $\alpha$, $\beta$ and $\gamma$ denoting the numbers of dots on the three lines). The diagrams of the remaining three rows are reducible. The name of each bubble diagram is a subscripted letter. The letter identifies the diagram of fig.~\ref{fig:6} obtained by cutting. The subscript distinguishes the bubble diagram from equivalent ones in figs.~\ref{fig:7}--\ref{fig:8}, obtained by different choices of origin.}%
	\label{fig:7}%
\end{figure}
\begin{figure}
	\centering
	\includegraphics[width=0.7\textwidth]{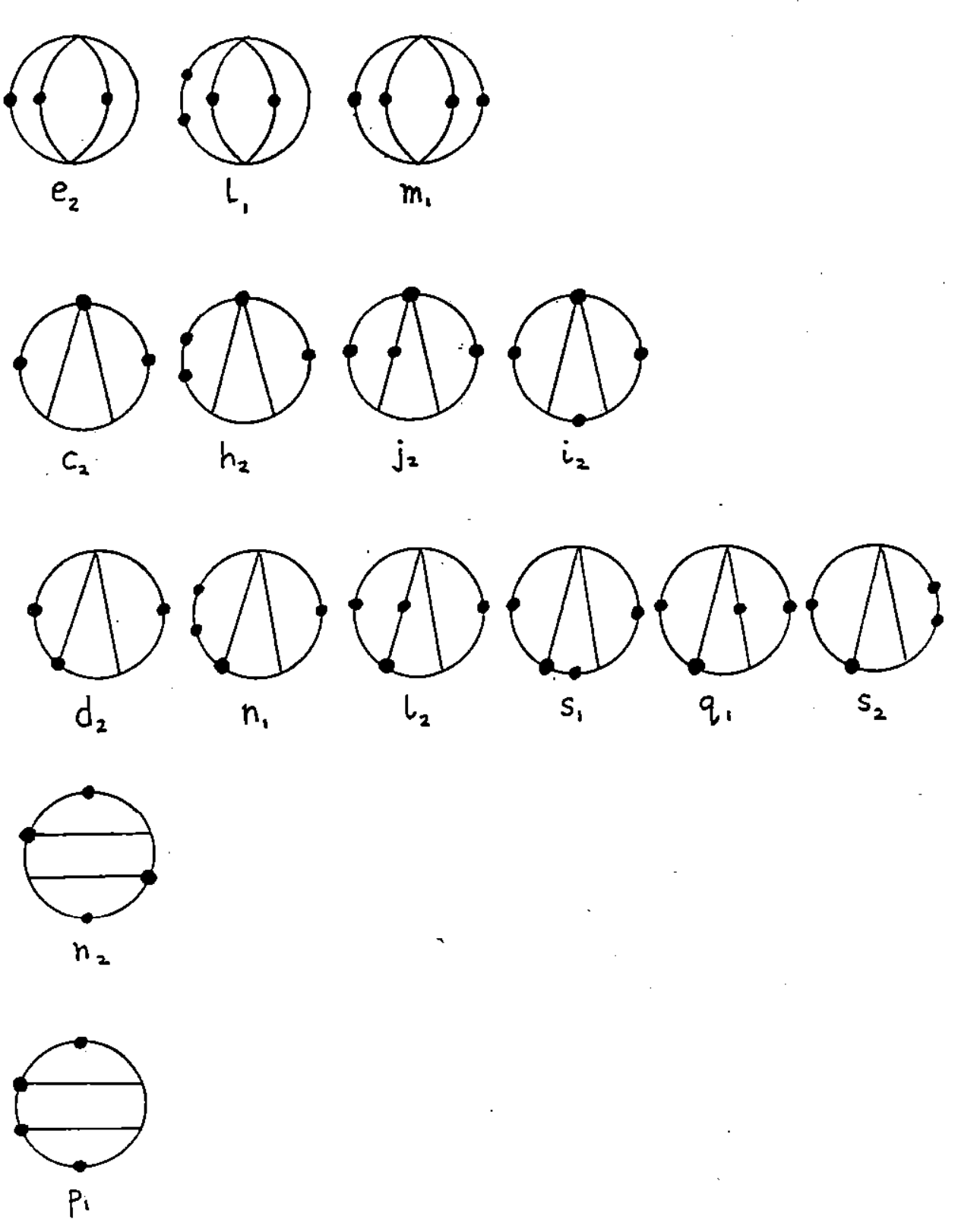}
	\caption{Harder bubble diagrams. The conventions are as in fig.~\ref{fig:7}. Conformal invariance reveals that only $\ell_1$, $m_1$, $n_1$ and $p_1$ need be evaluated, corresponding to the fourth row of fig.~\ref{fig:6}. This is done in subsections~\ref{sec:6.1} to \ref{sec:6.3}.}%
	\label{fig:8}%
\end{figure}
\begin{figure}
	\centering
	\includegraphics[width=0.7\textwidth]{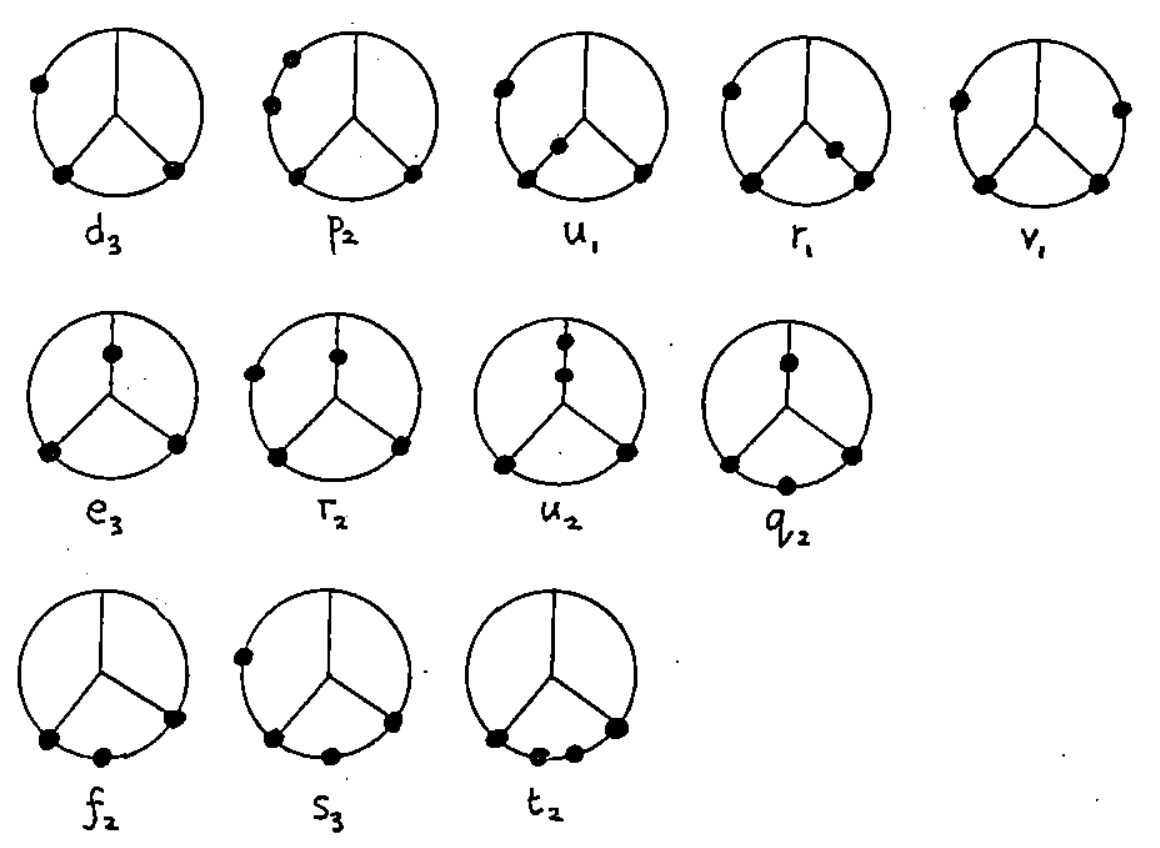}
	\caption{Superficially impossible bubble diagrams. The conventions are as in fig.~\ref{fig:7}. Conformal invariance reveals that none of these diagrams need be evaluated. This is fortunate, since GPXT \cite{2} would lead to six-fold sums, weighted by squares of $6$-$j$ co-efficients \cite{6}.}%
	\label{fig:9}%
\end{figure}

As thoroughly explained in ref.~\cite{2}, the efficiency of GPXT relies on a judicious choice of origin in $x$-space. For a two-point diagram there is a price to pay for not choosing one of the two external blobs as origin, since the failure to do so will lead to \emph{two} infinite series of Bessel functions in the expansion of $\exp(ik(x_1-x_2))$ in terms of Chebyshev polynomials with arguments $\hat{k}\cdot \hat{x}_1$ and $\hat{k}\cdot \hat{x}_2$.
On the other hand choosing a blob as origin has its drawbacks, since that may not be the best way of reducing the numbers of summations of the type \eqref{eq:2.14}.
No such dilemma faces us: there is no external momentum for the bubble diagrams and hence no Bessel function can appear. In such a situation the best choice of origin is usually clear: a vertex with as many lines as possible. Since each bubble diagram with at least $4$ loops has at least one $\phi^4$ (or higher order) vertex, it is clear that by choosing such a vertex as origin we can reduce even a $6$-loop bubble diagram to a three-loop angular diagram. Hence the method of exhaustion is to draw all angular diagrams with three loops or less and convert them into bubble diagrams by sprinkling them with dots to indicate points joined to the origin (which need not appear). This is a very compact way to represent a bubble diagram but it is not unique, since the same bubble diagram may be represented in different ways by different choices of origin.
Thus to generate fig.~\ref{fig:6} we first generated figs.~\ref{fig:7}--\ref{fig:9} by drawing angular diagrams up to three loops, furnished with the minimum numbers of dots to ensure only logarithmic divergence in the corresponding bubble diagrams. Dots were then added in all possible ways in each row, up to the maximum allowed by six-loop bubble diagrams, since no additional dot can create a subdivergence.
Then one has to identify equivalent bubble diagrams. This identification is notated by the device of labelling by the same letter, with a different numerical subscript.
(The reader might like to check a couple of identifications to get a feel for what is involved.)
This method appears to be fairly foolproof. The only danger is that one misses out a row or includes one with subdivergences. It is hoped that no row has been omitted. Certainly none has subdivergences. The only real subtlety is that the diagrams $n_2$ and $p_1$ of the last two rows of fig.~\ref{fig:8} are allowed, whilst the third possibility of putting the two vertex dots at the top leads to a subdivergence.

\subsection{Two-loop angular diagrams}
\label{sec:4.4}

It should be clear by now that diagrams $a$, $b$, $c$ and $g$ are now known, since the first row of fig.~\ref{fig:7} gives
\begin{equation}
	I_{\ell} = \binom{2\ell}{\ell} \zeta(2\ell-1);
	\quad \ell=2,3,4,5
	\label{eq:4.2}%
\end{equation}
for the coefficient of divergence of the $(\ell+1)$-loop bubble diagram, corresponding to the angular diagram with $\ell+1$ dots. This was, in essence, what prompted us to derive the result \eqref{eq:2.17} of subsection~\ref{sec:2.3}, since it was clear that the derivation of \eqref{eq:4.2} by dimensional regularization \cite{4} ought to be replaceable by a truly four-dimensional method based on GPXT.
To achieve this we needed $P_{\ell}(x,x)$ for eq.~\eqref{eq:2.20}. This led to finding $P_{\ell}(x_1,x_2)$ and now enables us to write a simple formula which generates, in principle, the second row of fig.~\ref{fig:7} (and continues for all loops).
Let $G(\alpha,\beta,\gamma)$ denote the value of any finite diagram obtained by cutting a line or vertex of an irreducible bubble diagram corresponding to a two-loop angular diagram with $\alpha$, $\beta$ and $\gamma$ dots on its three equivalent lines and one dot on a vertex.
Thus diagram $j$ has the value $G(2,1,1)$, since those are the numbers of dots on the lines of diagram $j_1$.
Then, following the same method of equating coefficients of divergence that was used in subsection~\ref{sec:2.3}, we have
\begin{equation}
	\pi^{-2} \int \frac{\dd[4]k}{k^4} G(\alpha,\beta,\gamma)
	= \pi^{-4} \int_{x_1^2>x_2^2} \dd[4] x_1 \dd[4] x_2
	\left[ \frac{1}{x_1^2} + \frac{1}{x_2^2} \right]
	P_{\alpha}(x_1,x_2) P_{\beta}(x_1,x_2) P_{\gamma}(x_1,x_2)
	\label{eq:4.3}%
\end{equation}
where we have first symmetrized on the points $x_{1,2}$ and then restricted to $x_1^2>x_2^2$. Using the expansion \eqref{eq:2.17} three times we can easily do the angular integration
\begin{equation}
	\int \dd \underline{\hat{x}}_1 \dd \underline{\hat{x}}_2
	C_{\ell}(\underline{\hat{x}}_1 \cdot \underline{\hat{x}}_2)
	C_{m}(\underline{\hat{x}}_1 \cdot \underline{\hat{x}}_2)
	C_{n}(\underline{\hat{x}}_1 \cdot \underline{\hat{x}}_2)
	= \triangle(\ell,m,n)
	\label{eq:4.4}%
\end{equation}
where $\ell$, $m$, $n$ are the positive integers corresponding to $n$ in \eqref{eq:2.17}. The symmetric function $\triangle(\ell,m,n)$ is that which appears in expressing a product of Chebyshev polynomials as a sum \cite{2,5,6}
\begin{equation}
	C_{\ell}(x) C_m(x)
	= \sum_{n} \triangle(\ell,m,n) C_n(x)
	\label{eq:4.5}%
\end{equation}
It is a sort of `triangular delta function' and is equal to $1$ or $0$ according as whether
\begin{equation*}
	g \equiv (\ell+m+n+1)/2
\end{equation*}
is or is not an integer satisfying
\begin{equation*}
	g > \Max(\ell,m,n)
\end{equation*}
The integration over $x_2^2<x_1^2$ must then yield a term proportional to $1/x_1^4$ to match the momentum-space divergence on the l.h.s.\ of \eqref{eq:4.3}. Equating divergences we find
\begin{multline}
	G(\alpha,\beta,\gamma)
	= \sum_{i,j,k} C(\alpha,i) C(\beta,j) C(\gamma,k) (i+j+k)! 
	\\
	\times S_+(2\alpha-i, 2\beta-j, 2\gamma-k, i+j+k+1)
	\label{eq:4.6}%
\end{multline}
where the coefficients $C$ are given by eq.~\eqref{eq:2.19} and 
\begin{equation}
	S_{\pm}(a,b,c,N)
	\equiv \sum_{\ell,m,n} \frac{\triangle(\ell,m,n)}{\ell^a m^b n^c} 
	\left[ \left(\frac{1}{g-1}\right)^N \pm \left(\frac{1}{g}\right)^N \right]
	\label{eq:4.7}%
\end{equation}
The importance of $S_-$ will soon become apparent. The remaining problem is to reduce the constrained triple sums $S_+$ to their simplest forms, which is where the miracles begin.

\subsection{From triple sums to $\zeta$ functions}
\label{sec:4.5}

First we use the recurrence relations
\begin{equation}\begin{split}
	S_{\pm}(a,b,c,N)
	&= S_{\mp}(a-1,b,c,N) + S_{\mp}(a,b-1,c,N) + S_{\mp}(a,b,c-1,N)
	\\& - 2 S_{\mp}(a,b,c,N-1)
	\label{eq:4.8}%
\end{split}\end{equation}
to reduce all the $S_+$ functions to $S$ functions with one vanishing argument. But $S_-$ vanishes when $N=0$ and, \emph{mirabile dictu}, we find that in all cases up to $7$ loops (the maximum so far investigated) the terms involving $S_+(a,b,c,0)$ cancel completely, after taking account of the symmetry in $a$, $b$, $c$: yet another unexpected and, as yet, generally unproved blessing. We thus need never struggle with the apparently intractable sums
\begin{equation}
	S_+(a,b,c,0)
	= 2 \sum_{\ell,m,n} \frac{\triangle(\ell,m,n)}{\ell^a m^b n^c}
	\label{eq:4.9}%
\end{equation}
and instead need only work on the much more yielding sums
\begin{equation}
	T_{\pm} (a,b,N)
	= \sum_{\ell+m \geq g > \ell,m} \frac{1}{\ell^a m^b} \left[ \left( \frac{1}{g-1} \right)^N \pm \left( \frac{1}{g} \right)^N \right]
	\label{eq:4.10}%
\end{equation}
which are trivially given by double sums in the case of $T_-$.

To organise an analysis of constrained triple sums it is useful to relate everything to
\begin{subequations}\label{eq:4.11}%
\begin{align}
	F(a,b,c) &\equiv \sum_{i>j>k>0} \frac{1}{i^a j^b k^c}
	\label{eq:4.11a}%
	\\
	F(a,b) &\equiv \sum_{i>j>0}\frac{1}{i^a j^b}
	\label{eq:4.11b}%
\end{align}
\end{subequations}
To do this we split $T_+$ of eq.~\eqref{eq:4.10} into four separate sums as follows:
\begin{equation*}
	\sum_{\ell+m\geq g>\ell,m}
	= \sum_{g>\ell>m} + \sum_{g>m>\ell} + \sum_{g>\ell=m} - \sum_{g>\ell+m}
\end{equation*}
The first three terms are of the desired form; the last one leads to triple sums of the form
\begin{equation}
	\sum_{\ell,m,n} \frac{1}{\ell^a m^b (\ell+m+n)^N}
	\label{eq:4.12}%
\end{equation}
which can be reduced to sums of the desired form by repeatedly using
\begin{equation}
	\frac{1}{\ell m} = \frac{1}{\ell+m}\left[ \frac{1}{\ell} +\frac{1}{m} \right]
	\label{eq:4.13}%
\end{equation}
Having thus arrived at sums like \eqref{eq:4.11} without great difficulty we searched the mathematical literature for possible reductions to $\zeta$ functions, and were disappointed to find only the simple results
\begin{equation*}
	F(2,1) = \zeta(3);\quad F(3,1) = \zeta(4)/4
\end{equation*}
in Lewin's book \cite{11} on polylogarithms, attributed to a work by Nielsen in 1909. Left to our own devices we succeeded in evaluating \emph{all} of the double and triple sums needed to $5$ loops, in terms of $\zeta$ functions. The trick is elementary in the extreme and we sketch it here to prevent others from spending as long discovering as we did. Consider first the double sums \eqref{eq:4.11b}. By use of
\begin{equation}
	\sum_i \sum_j = \sum_{i>j} + \sum_{j>i} + \sum_{j=i}
	\label{eq:4.14}%
\end{equation}
one trivially finds that
\begin{equation}
	\zeta(a) \zeta(b) = F(a,b) + F(b,a) + \zeta(a+b)
	\label{eq:4.15}%
\end{equation}
enabling one to eliminate $F(a,b)$ when $b \geq a$. On the other hand one can use \eqref{eq:4.13} to express
\begin{equation*}
	\zeta(a) \zeta(b) = \sum_{\ell,m}\frac{1}{\ell^a m^b}
\end{equation*}
in terms of combinations of 
\begin{equation*}
	F(a,b) = \sum_{\ell,m} \frac{1}{(\ell+m)^a \ell^b}
	= \sum_{\ell,m} \frac{1}{(\ell+m)^a m^b}
\end{equation*}
The result is a system of linear relations between $F(a,b)$ functions with fixed $a+b$. The system completely determines the sums if $a+b$ is odd and leaves just one undetermined if $a+b$ is even and greater than $6$.
(By a principle of conservation of information one might expect that when $a+b$ is even the price of not knowing one $F(a,b)$ buys one something. In fact, it tells one the Bernoulli number $B_{a+b}$.)

The principles of finding $F(a,b,c)$ of eq.~\eqref{eq:4.11a} are similar, but the work is more demanding. The decomposition of factorizable triple sums, comparable to \eqref{eq:4.14}, involves six different $F(a,b,c)$ and three different $F(a,b)$, as well as $\zeta$.
It turns out that when $a$, $b$, $c$ are all different only two of the six permutations of the arguments of $F(a,b,c)$ need be kept, the other four permutations being determined by these two.
With two equal arguments only one permutation survives, and with three equal arguments none survives. Now that the set of triple sums \eqref{eq:4.11a} has been reduced by relations analogous to \eqref{eq:4.15} one needs a trick analogous to \eqref{eq:4.13}. A suitable one is to use
\begin{equation}
	\frac{1}{n(\ell+m)}
	= \frac{1}{\ell+m+n} \left[ \frac{1}{n} + \frac{1}{\ell+m} \right]
	\label{eq:4.16}%
\end{equation}
to turn the factorizable sums
\begin{equation*}
	\sum_{\ell,m,n} \frac{1}{\ell^a n^b (\ell+m)^c}
	=\zeta(b) F(c,a)
\end{equation*}
into ones of the form \eqref{eq:4.11a} or of the form \eqref{eq:4.12}.
But one has already converted sums \eqref{eq:4.12} into sums \eqref{eq:4.11}.
Thus a system of linear relations, preserving the value of $a+b+c$, results.
Remarkably this fixes all the $F(a,b,c)$ with $a+b+c\leq 9$, \emph{precisely} what one needs for $5$-loop calculations.
And there is no danger that the first unknown $F(a,b)$ will appear, since it has $a+b=8$ and there is no such thing as $\zeta(1)$ by which to multiply it.
Once more guardian angels have been at work and everything is given by $\zeta$ functions. Not surprisingly they have also ensured that when one combines a result such as
\begin{equation}
	F(5,3,1) = \frac{845}{24} \zeta(9) - 17\zeta(7) \zeta(2) -\frac{3}{4}\zeta(6) \zeta(3)
	-\frac{23}{4} \zeta(5) \zeta(4) + \frac{1}{6} \big[ \zeta(3) \big]^3
	\label{eq:4.17}%
\end{equation}
with all the other ones required by eqs.~\eqref{eq:4.6} and \eqref{eq:4.7}, the $\zeta$ functions with even arguments vanish in a puff of smoke leaving diagram $j$, for example, as
\begin{equation*}
	G(2,1,1)
	= 108 \zeta(5) \zeta(3) + \frac{189}{2} \zeta(7) - 36 \big[ \zeta(3) \big]^2
\end{equation*}
One's immediate reaction is that there ought to be a more direct way to such a result. This is surely true. Yet it seems that beyond $5$ loops the age of miracles ends, or at least abates.
For example both of the six-loop diagrams $G(3,2,0)$ and $G(4,1,0)$ involve $F(8,2,1)$, which has resisted all attempts at reduction to $\zeta$ functions. But then on the other hand one finds that
\begin{equation}
	G(3,2,0)+4 G(4,1,0)
	= \frac{9163}{3} \zeta(11) + 160 \zeta(5) \big[ \zeta(3) \big]^2
	\label{eq:4.18}%
\end{equation}
so perhaps some \emph{combinations} of diagrams can be obtained directly by expanding $\Gamma$ functions or their derivatives.

\subsection{What remains}
\label{sec:4.6}

The attentive reader may have spotted the principle of organisation of figs.~\ref{fig:7}--\ref{fig:9}.
Fig.~\ref{fig:7} gives all the diagrams that are now known, since the one-loop angular diagrams of the first row were disposed of in section~\ref{sec:2} and the two-loop angular diagrams of the second row have now succumbed. The remaining diagrams of fig.~\ref{fig:7} are obviously reducible, giving
\begin{equation}
	f = a \times a
	\qquad
	t=k=a \times b
	\label{eq:4.19}%
\end{equation}
for the corresponding diagrams of fig.~\ref{fig:6}.

So what remains are those diagrams of fig.~\ref{fig:6} not represented in fig.~\ref{fig:7}. The division of these between fig.~\ref{fig:8} and fig.~\ref{fig:9} is simple: the former contains very difficult diagrams and those of the latter appear impossible. We shall see.

\section{Conformal invariance}
\label{sec:5}

There is a result of the previous analysis which is most intriguing: diagrams $e$ and $f$ are both equal to $[6\zeta(3)]^2$. This is no surprise for $f$, since diagram $f_1$ is clearly reducible. But diagram $e_1$ is a non-planar irreducible bubble diagram. Why should it, apparently the most difficult $5$-loop bubble diagram, be equal to the easiest one?
The answer is not to be discovered by looking at $e_1$ and $f_1$ in fig.~\ref{fig:7}, but rather at $e_3$ and $f_2$ in fig.~\ref{fig:9}, which give very foolish prescriptions for using GPXT to calculate $e$ and $f$, since each involves a six-fold sum of a complicated rational function of six integers weighted by a coefficient which Nickel succeeded in expressing as the \emph{square} of a \emph{three}-dimensional $6$-$j$ coefficient \cite{6}.
But the observation that $e_3=f_2$ is the key to a method to abolish fig.~\ref{fig:9} entirely. To change one diagram into the other, one can move either the dot on one of the six lines, or both dots on two of the four vertices. The second transformation is in fact effected very simply. One merely performs the inversion
\begin{equation}
	x_{\mu} \rightarrow x_{\mu}/x^2
	\label{eq:5.1}%
\end{equation}
Conformal invariance ensures that the coefficient of logarithmic divergence of a dimensionless massless bubble diagram will be unchanged. But the effect is to change an undotted cubic vertex into a dotted one (and vice versa); to leave an undotted quartic vertex unchanged; and to turn a dotted quartic vertex into something quite outside the province of scalar bubble diagrams.
This is because each line drawn in figs.~\ref{fig:7}--\ref{fig:9} carries a factor $(x_1^2 x_2^2)^{-1/2}$, irrespective of the number of dots which may lie on it. Eqs.~\eqref{eq:2.15} and \eqref{eq:2.17} show that, for all $\ell$,
\begin{equation*}
	(x_1^2 x_2^2)^{1/2} P_{\ell}(x_1,x_2)
\end{equation*}
is invariant under the transformation \eqref{eq:5.1}. So an undotted quartic vertex is unchanged, since it picks up four factors of $(x^2)^{-1/2}$ to give the invariant measure $\int \dd[4] x/x^4$. But a dotted cubic vertex gets $(x^2)^{-3/2}$ from its three lines and $(x^2)^{-1}$ from its dot, giving $\int \dd[4] x/(x^2)^{5/2}$, whilst its undotted cousin only gets $\int \dd[4] x/(x^2)^{3/2}$.
The transformation \eqref{eq:5.1} swaps roles.

Thus conformal invariance is a most generous friend, giving us the following relations between diagrams:
\begin{equation}\label{eq:5.2}\begin{split}
	n_1 &= s_2,\qquad \ell_2 = q_1,\qquad  u_1 = v_1, \\
	e_3 &= f_2,\qquad r_2 = s_3,\qquad     u_2 = t_2.
\end{split}\end{equation}
Inspection of fig.~\ref{fig:9} now reveals that \emph{none} of the $12$ diagrams it contains need be evaluated as a Mercedes angular diagram. If a diagram has a subscript greater $1$ we may ignore it, since there is an alternative. That leaves only $u_1$, $r_1$ and $v_1$. But eqs.~\eqref{eq:5.2} reveal that
\begin{equation*}
	v_1 = u_1 = u_2 = t_2 = t_1
\end{equation*}
which promotes $v$ and $u$ to the top category of diagrams already evaluated in fig.~\ref{fig:7} as reducible. We get
\begin{equation*}
	v=u=t=k=a \times b
\end{equation*}
from eq.~\eqref{eq:4.19}. That leaves only
\begin{equation*}
	r_1=r_2=s_3=s_2=n_1
\end{equation*}
which promotes $r$ from the dungeon of fig.~\ref{fig:9} to the ground level of fig.~\ref{fig:8}. Moreover one can isolate a minimal subset of the diagrams of fig.~\ref{fig:8} requiring evaluation. There is no need to bother with
\begin{equation*}
	q_1 = \ell_2 = \ell_1
	\qquad
	s_1 = s_2 = n_1
\end{equation*}
Ignoring subscripts greater than $1$ we are left with
\begin{equation*}
	\set{\ell_1, m_1, n_1, p_1}
\end{equation*}
That means that the first three rows of fig.~\ref{fig:6} are done and the last two are unnecessary. Only the fourth row remains: $17$ down and $4$ to go.

\section{Diagrams remaining}
\label{sec:6}

The first row of fig.~\ref{fig:8} gives a glimmer of hope for $\ell_1$ and $m_1$, since it also contains $e_2=e_1$, which is already known:
\begin{equation}
	e = f = a\times a = \left[ 6 \zeta(3) \right]^2
	\label{eq:6.1}%
\end{equation}
To tackle it we will need to evaluate quadruple sums weighted by
\begin{equation}
	\Box(a,b,c,d)
	\equiv \int \dd \underline{\hat{x}}_1 \dd \underline{\hat{x}}_2
	C_a (\underline{\hat{x}}_1 \cdot \underline{\hat{x}}_2)
	C_b (\underline{\hat{x}}_1 \cdot \underline{\hat{x}}_2)
	C_c (\underline{\hat{x}}_1 \cdot \underline{\hat{x}}_2)
	C_d (\underline{\hat{x}}_1 \cdot \underline{\hat{x}}_2)
	\label{eq:6.2}%
\end{equation}
The third row offers more hope for $n_1$, since it contains $d_2=d_1$, which is known, and this row does not look too different from the second row, \emph{all} of whose members were calculated in their previous incarnations in fig.~\ref{fig:7}. This also bodes well for $p_1$ in the final row, since its close cousin $n_2$, above it, is equal to $n_1$ of the third row. 
Accordingly we tackle the diagrams in the order $n_1$, $p_1$, $\ell_1$ and $m_1$, discovering a series of three more `miracles'.

\subsection{Another miracle}
\label{sec:6.1}

The strategy for evaluating $n_1$ should be clear. We take account of the dots on its lines by using the appropriate values of $\ell$ in eq.~\eqref{eq:2.17}.
We use eq.~\eqref{eq:4.5} twice, to dispose of the two subloops of the angular diagram with only two lines each.
The integration over $x_{1,2,3}$ is symmetrized and then restricted to $x_1^2>x_2^2>x_3^2$, the final integration over $x_1^2$ giving the trivial $\int \dd[4] x_1/x_1^4$ of logarithmic divergence.
The dotted vertex merely corresponds to dividing by the appropriate $x^2$ before symmetrization.
As a check on this procedure we perform the identical procedure for \emph{all} the diagrams in the second and third rows of fig.~\ref{fig:8}.
Using \texttt{REDUCE3} we easily generate the rational functions $R(a,b,c,d,e)$ in the resulting sums
\begin{equation}
	\sum_{a,b,c,d,e} \triangle(a,b,c) \triangle(c,d,e) R(a,b,c,d,e)
	\label{eq:6.3}%
\end{equation}
They are horrendous! Despairing of an analytical result we output $R$ as many lines of \texttt{FORTRAN} and resort to a truncated numerical sum, obtaining good agreement with all known relations.
The only hope now is that there is a simple linear relation between the unknown diagram $n_1$ and some of its known cousins in rows $2$ and $3$ of fig.~\ref{fig:8}. Indeed there is. We have verified to $9$ significant figures that
\begin{equation*}
	n = h - i/3
\end{equation*}
Unfortunately no such relation for $\ell$ emerges from rows $2$ and $3$ of fig.~\ref{fig:8}, despite intensive numerical investigation of diagram $\ell_2$.

\subsection{And another}
\label{sec:6.2}

The procedure for $p_1$ is little different, but the computer takes a little longer because now there are $24$ symmetrizations of four points, and three non-trivial radial integrations. This time the miracle emerges analytically: the summands for $p_1$ and $i_2$ in \eqref{eq:6.3} are revealed by \texttt{REDUCE3} to be identical. Thus
\begin{equation*}
	p = i
\end{equation*}
Note that the existence of a sixth line in $p_1$ is deceptive. The sum is still only five-fold, since the orthogonality of Chebyshev polynomials sets the integers of summation equal on the two vertical lines of $n_2$ and $p_1$.

\subsection{And now the last?}
\label{sec:6.3}

The `square delta function' \eqref{eq:6.2} has the following properties: it vanishes unless
\begin{equation}
	h \equiv (a+b+c+d)/2
	\label{eq:6.4}%
\end{equation}
is an integer satisfying
\begin{equation}
	h > \Max(a,b,c,d)
	\label{eq:6.5}%
\end{equation}
in which case
\begin{equation}
	\Box = \Min(a,b,c,d,h-a,h-b,h-c,h-d)
	\label{eq:6.6}%
\end{equation}
This follows from the convolution of two `triangular delta functions':
\begin{equation}
	\Box(a,b,c,d) = \sum_e \triangle(a,b,e) \triangle(e,c,d)
	\label{eq:6.7}%
\end{equation}
The sums we require for $e_2$, $\ell_1$, $m_1$ are of the form
\begin{equation*}
	\sum_{a,b,c,d} \frac{\Box(a,b,c,d)}{a^{\alpha} b^{\beta} c^{\gamma} d^{\delta} h^N}
\end{equation*}
By using partial fractions we can reduce these to sums of terms with $\delta=0$, which lead to sums of the form
\begin{equation*}
	\sum_{h>a,b,c} \frac{\mu(a,b,c;h)}{a^{\alpha}b^{\beta}c^{\gamma} h^N}
\end{equation*}
with the measure
\begin{equation}\begin{split}
	\mu(a,b,c;h)
	&\equiv \abs{2h-a-b-c} + \abs{h-a-b-c}
	\\&
	-\abs{h-a-b} - \abs{h-b-c} - \abs{h-c-a}
	\label{eq:6.8}%
\end{split}\end{equation}
resulting from eqs.~\eqref{eq:6.4}--\eqref{eq:6.6}. Note that, by construction, $\mu$ vanishes for $h>a,b,c$ unless
\begin{equation}
	a+b+c>h>(a+b+c)/2
	\label{6.9}%
\end{equation}
corresponding to the requirement that
\begin{equation}
	h>d>0
	\label{eq:6.10}%
\end{equation}
In fact $\mu/2$ is just a convenient way of expressing $\Box$ after eliminating $d$.

Sums involving the first term in the measure \eqref{eq:6.8} have proved very resistant to further simplification, whereas we believe that one might, in principle, relate all the others to quadruple sums of the form
\begin{equation}
	F(a,b,c,d)
	\equiv \sum_{i>j>k>\ell>0} \frac{1}{i^a j^b k^c \ell^d}
	\label{eq:6.11}%
\end{equation}
and replay the game of subsection~\ref{sec:4.5} at a level of complexity which would probably stretch both \texttt{REDUCE3} and the author past breaking point. A final miracle is needed, and one must be at hand, since after all we know that
\begin{equation*}
	e = a \times a = \left[ 6 \zeta(3) \right]^2
\end{equation*}
The miracle for diagram $e_2$ is demystified by the realization that the rational function $E$ of
\begin{equation*}
	e = \sum_{h>a,b,c} \mu(a,b,c;h) E(a,b,c;h)
\end{equation*}
has the property that when one multiplies it by $(2h-a-b-c)$ one generates terms with only \emph{three} reciprocal powers of $a$, $b$, $c$ or $h$, instead of the four that bar further progress for the generic contribution of the first term of \eqref{eq:6.8}.
This suggests that if one can find a combination of diagrams $\ell_1$ and $m_1$ for which the same miracle happens, one might get a multiple of $a\times b$, since a multiple of $\zeta(3) \zeta(5)$ is the only simple expected result with a summand involving reciprocal powers which sum to $8$. It is not hard to find such a combination analytically and then to verify numerically the relation
\begin{equation}
	16 \ell + m = 12 a \times b = 1440 \zeta(3) \zeta(5)
	\label{eq:6.12}%
\end{equation}
to $9$ significant figures.

That leaves just one diagram, $m$, outside the fold. The precise quadruple sum is
\begin{equation}
	m = 16J,\quad
	J \equiv \sum_{h>a,b,c} \frac{\mu(a,b,c;h)}{a^2b^2c^2h^3} \left\{ 2 + 9 \frac{a}{h} + 18 \frac{ab}{h^2} + 15 \frac{abc}{h^3} \right\}
	\label{eq:6.13}%
\end{equation}
with $\mu$ given by eq.~\eqref{eq:6.8}.

At this point we believe that the age of miracles is ended and present the full table of results.

\subsection{Results}
\label{sec:6.4}

The results for all the diagrams of fig.~\ref{fig:6} are given in table~\ref{tab:3}.
\begin{table}
	\centering
	\caption{Values of diagrams of fig.~\ref{fig:6}.}%
	\label{tab:3}%
	\begin{tabular}{>{$}c<{$}>{$}c<{$}>{$}r<{$}>{$}l<{$}}
		\text{diagram} & \text{loops} & \text{mult.} & \text{value} \\
	\toprule
		a & 2 & 1 & 6 \zeta(3) \\
	\midrule
		b & 3 & 4 & 20 \zeta(5) \\
	\midrule
		c & 4 & 4 & 70 \zeta(7) \\
		d & 4 & 9 & 441 \zeta(7)/8 \\
		e & 4 & 4 & 36[\zeta(3)]^2 \\
		f & 4 & 4 & \text{as for $e$}\\
	\midrule
		g & 5 & 8 & 252 \zeta(9) \\
		h & 5 & 25& 1567 \zeta(9)/9 + 8[\zeta(3)]^3 \\
		i & 5 & 14& 168 \zeta(9) \\
		j & 5 & 18& 108\zeta(5)\zeta(3) + 189\zeta(7)/2 - 36[\zeta(3)]^2\\
		k & 5 & 14& 120 \zeta(5)\zeta(3)\\
		\ell&5& 10& 90\zeta(5)\zeta(3) - J \\
		m & 5 & 2 & 16J = 71.506081796562 \\
		n & 5 & 12 & 1063\zeta(9)/9 + 8[\zeta(3)]^3 \\
		p & 5 & 13 & \text{as for $i$} \\
		q & 5 & 10 & \text{as for $\ell$} \\
		r & 5 & 13 & \text{as for $n$} \\
		s & 5 & 21 & \text{as for $n$} \\
		t & 5 & 11 & \text{as for $k$} \\
		u & 5 & 13 & \text{as for $k$} \\
		v & 5 & 6  & \text{as for $k$} \\
	\bottomrule
	\end{tabular}
\end{table}

For each diagram we give the multiplicity of equivalent diagrams, obtained by gluing and cutting. Thus the $216$ diagrams are reduced to the $21$ diagrams of fig.~\ref{fig:6}, only $13$ of which are numerically distinct in table~\ref{tab:3}.
The $13$ different values are expressed rationally in terms of just $5$ numbers: $\zeta(3)$, $\zeta(5)$, $\zeta(7)$, $\zeta(9)$ and
\begin{equation}
	J=4.4691301122851
	\label{eq:6.14}%
\end{equation}
defined by eqs.~\eqref{eq:6.8} and \eqref{eq:6.13}.

Note that a numerical result is given for $J$ to $14$ significant figures. We were surprised to be able to achieve this for a four-fold sum with complicated constraints, and hope that the estimated error of $10^{-14}$, reflecting the stability on different computations, is realistic.
The method was based on the analytical result that the sum over $a,b,c<h$ in \eqref{eq:6.13} produces a summand over $h$ which behaves as
\begin{equation*}
	S(h) \sim \frac{(\log h)^2}{h^5}
	\quad\text{as}\quad
	h \rightarrow \infty
\end{equation*}
To evaluate
\begin{equation*}
	J = \sum_{h=2}^{\infty} S(h)
\end{equation*}
we truncated the sum at $h=100$, performing $10^8$ evaluations of the summand of \eqref{eq:6.13}.
Whilst accumulating the values of $S(2)$ to $S(100)$ we sought constantly to improve the truncated sum by a method of finite differences equivalent to the approximation
\begin{equation*}
	S(h) \simeq \sum_{n=0}^{2} \sum_{m=5}^{10} C(n,m) \frac{(\log h)^n}{h^m}
\end{equation*}
achieved by taking $18$ steps of successive improvement by finite differences.
The stability was remarkable, the quoted value \eqref{eq:6.14} being attained at the 13th step of improvement with $h \leq 100$ and at the 18th step with $h\leq 70$, with consistent results in between.
We thought this effort worthwhile because there still remains the possibility of a simpler result for $m=16J$.
However, our best `random' guess
\begin{equation*}
	\frac{13}{3} F(6,2) + \frac{25}{12} \big[ \zeta(3) \big]^2 \zeta(2) + \frac{160}{3} \zeta(5) \zeta(3)
\end{equation*}
fails to reproduce the computed value of $m$ beyond the tenth significant figure.

If, as we strongly suspect, $J$ is a genuinely new transcendental number, it might show up in $6$-loop beta functions. Such a tentative conclusion is a little melancholy, since it undermines the pious hope that ultimately someone will discover a way of achieving the results reported here with far greater economy of effort and far less guesswork.
But \emph{if} there is a stumbling block at the level of six-loop bubble diagrams, $m_1$ has all the right credentials to be it: it is totally symmetric in its twelve lines, its three $\phi^4$ vertices, and its four $\phi^3$ vertices, and is highly non-planar: the classic marks of a very tough customer.
Even so, past experience discourages me from betting heavily against new miracles, even here.

\section{Conclusions, puzzles and appeals}
\label{sec:7}

It may help to collect these from previous sections.

\subsection{Conclusions}
\label{sec:7.1}

The diagrams of fig.~\ref{fig:1} can be easily evaluated by using eq.~\eqref{eq:2.10} for arbitrary dimensions, loops and exponents.
The four-dimensional result \eqref{eq:2.12} is more transparently derived by GPXT, using the `$\ell$-fold propagator' of eqs.~\eqref{eq:2.16}, \eqref{eq:2.17} and \eqref{eq:2.19}.
The diagrams of fig.~\ref{fig:4} present greater difficulty, even after reduction to the first.
In general one needs to evaluate the single sums of eq.~\eqref{eq:3.9}.
But for diagrams contributing to $5$-loop beta functions eq.~\eqref{eq:3.26} should suffice, and for higher orders eqs.~\eqref{eq:3.30} and \eqref{eq:3.31} enable one to dispose of modestly divergent diagrams, reducible to fig.~\ref{fig:4}.
Even so many diagrams are not reducible to figs.~\ref{fig:1} and \ref{fig:4}.
As a partial antidote to this restriction we offer the results of table~\ref{tab:3} for all the diagrams of fig.~\ref{fig:6}, evaluated via the angular diagrams of figs.~\ref{fig:7} and \ref{fig:8}.
Because of conformal invariance one need not tackle the much more difficult angular diagrams of fig.~\ref{fig:9}.

\subsection{Puzzles}
\label{sec:7.2}

The author is unable to answer the following questions:
\begin{enumerate}[(a)]
	\item
		Is there a master function for fig.~\ref{fig:4}, analogous to, but more complicated than, the master function $P$ of eq.~\eqref{eq:2.10} for fig.~\ref{fig:1}? If so, it might involve square roots and/or derivatives of gamma functions.

	\item
		Can the reflection symmetry of eq.~\eqref{eq:3.15} be derived without going through the complicated steps of eqs.~\eqref{eq:3.2}--\eqref{eq:3.14}? If so, it might turn out to be a member of a large class of reflection symmetries.

	\item
		Can the relation between figs.~\ref{fig:1} and \ref{fig:4}, generalizing eqs.~\eqref{eq:3.28} and \eqref{eq:3.29}, be derived without recourse to recurrence relations? If so, it might turn out to be a member of a large class of relations between families of diagrams of different topologies.

	\item
		Why do the apparently intractable sums \eqref{eq:4.9} disappear from \eqref{eq:4.6} when using \eqref{eq:4.8}? Is this true for all loops, rather than just for the $7$ so far investigated?

	\item
	Are the apparently intractable cases of \eqref{eq:4.11}, such as $F(6,2)$ and $F(8,2,1)$, truly irreducible to products of $\zeta$ functions? If so, what is the rule for what is reducible and why does it allow \emph{all} $5$ loop diagrams of the type \eqref{eq:4.6} to be reduced?

	\item
		Why has $\pi^2$ never emerged, via $\zeta$ functions with even arguments, when a diagram is reducible to $\zeta$ functions?

	\item
		What is the origin of the numerically discovered miracle $n=h-i/3$ of subsection~\ref{sec:6.1}?

	\item
		What is the origin of the analytically discovered miracle $p=i$ of subsection~\ref{sec:6.2}?

	\item
		Can one obtain, analytically, the results
		\begin{equation*}
			e = a \times a
			\qquad 16\ell + m = 12 a\times b
		\end{equation*}
		from row $1$ of fig.~\ref{fig:8}, without tackling the frightening sums \eqref{eq:6.11}?

	\item
		Is $m$ the only intractable diagram at the level of $6$-loop beta functions?

	\item
		May Racah coefficients be similarly avoided at higher loops?
\end{enumerate}
In addition to answers to these questions, the following are also lacking:
\begin{enumerate}[(a)]\setcounter{enumi}{11}
	\item
	An analytic proof of eqs.~\eqref{eq:3.14} and \eqref{eq:3.15}, which have as yet received only exhaustive numerical validation in the generic case of non-integer $\alpha$, $\beta$, $\gamma$ and $n$.

	\item
		A systematic algorithm for extending expansion \eqref{eq:3.26}.

	\item
		A proof of \eqref{eq:3.31} for all loops.
\end{enumerate}

\subsection{Appeals for further work}
\label{sec:7.3}

The list (a)--(n) above is as long and diverse as the list of results of table~\ref{tab:3}. This is significant, since the former gives the unsolved puzzles which seem to me to be more important than the latter's values.
To ingenious readers I appeal for solutions. But there is a larger appeal: to find a route to such results that illuminates perturbation theory, rather than compounds the mystery of calculational complexity and simplicity of the final answer. When that challenge is met, much else may become possible.

\subsection*{Acknowledgements}
I thank Varouzhan Baluni, David Barfoot, Paul Clark and Sotos Generalis for advice and encouragement at crucial stages of a long and arduous investigation.
I am most grateful for the hospitality and financial assistance of the CERN theory division, without which the investigations of sections \ref{sec:4}--\ref{sec:6} would not have been undertaken, let alone completed.

\bibliography{refs}
\bibliographystyle{JHEPdoi}

\end{document}